\documentclass[aps,prl,twocolumn,preprintnumbers,groupedaddress]{revtex4-2}
\usepackage{amsmath,amssymb,slashed}
\usepackage{axodraw2}
\usepackage{graphicx,xcolor}
\usepackage[colorlinks=true, allcolors=blue]{hyperref}

\newcommand{\msbar}{\overline{\text{MS}}}
\newcommand{\swft}{s^4_W}
\newcommand{\swsq}{s^2_W}
\newcommand{\cwsq}{c^2_W}

\newcommand{\ue}[1]{u_{#1}}
\newcommand{\je}[1]{J_{#1}}
\newcommand{\lmz}{l^{\mu}_z}
\newcommand{\lwz}{l^w_z}
\newcommand{\lchiO}{l^{\chi}_0}
\newcommand{\lmO}{l^{\mu}_0}
\newcommand{\HBchiPT}{HB$\chi$PT}
\newcommand{\LLs}{\text{LL}_s}
\newcommand{\NLLs}{\text{NLL}_s}

\begin{document} 

\title{Beyond Leading Logarithms in $g_V$: The Semileptonic Weak Hamiltonian at $\mathcal{O}(\alpha\,\alpha_s^2)$}

\preprint{P3H-25-087, TTP25-041}

\author{M.~Gorbahn}
\email{Martin.Gorbahn@liverpool.ac.uk}
\affiliation{Department of Mathematical Sciences, University of Liverpool, Liverpool L69 3BX, United Kingdom}

\author{F.~Moretti}
\email{Francesco.Moretti@kit.edu}
\affiliation{Institut f{\"u}r Theoretische Teilchenphysik, Karlsruhe Institute of Technology (KIT), Wolfgang-Gaede Stra\ss{}e 1, 76131 Karlsruhe, Germany,\\ 
and Department of Mathematical Sciences, University of Liverpool, Liverpool L69 3BX, United Kingdom}

\author{S.~Jäger}
\email{s.jaeger@sussex.ac.uk}
\affiliation{Department of Physics and Astronomy, University of Sussex, Falmer, Brighton BN1 9QH, United Kingdom}

\begin{abstract}
We present the first next-to-leading-logarithmic QCD analysis of the electromagnetic corrections to the semileptonic weak Hamiltonian, including the mixed $\mathcal{O}(\alpha\,\alpha_s^2)$ corrections to the vector coupling $g_V$. 
The analysis combines the evaluation of three-loop anomalous dimensions and two-loop matching corrections with a consistent factorization of short-distance QCD effects. 
The latter is implemented through a scheme change based on a $d$-dimensional operator product expansion performed inside the loop integrals. 
The resulting renormalization-group--improved expression for the radiative correction $\Delta^V_R = 2.436(16)\%$ can be systematically refined using input from lattice QCD and perturbation theory and improves the consistency of first-row CKM unitarity tests.
\end{abstract}

\maketitle


Within the Standard Model, semileptonic decay rates offer a clean determination of the Cabibbo–Kobayashi–Maskawa (CKM) matrix elements.
High-precision measurements of these decays are central to current tests of flavor physics and, in particular, to tests of first-row CKM unitarity, $|V_{ud}|^2 + |V_{us}|^2 + |V_{ub}|^2 = 1$.
Using current world averages \cite{ParticleDataGroup:2024cfk}, this relation appears to be violated at the level of 2--3$\sigma$.
Although such a deviation could point to physics beyond the Standard Model~\cite{Belfatto:2019swo,Coutinho:2019aiy,Belfatto:2021jhf,Cheung:2020vqm,Branco:2021vhs,Crivellin:2020ebi,Kirk:2020wdk,Crivellin:2021njn,Crivellin:2020lzu,Capdevila:2020rrl,Crivellin:2020lzu,Crivellin:2021sff,Crivellin:2020oup,Marzocca:2021azj,Cirigliano:2009wk,Alok:2021ydy,Cirigliano:2023nol,Dawid:2024wmp}, we strengthen the theoretical framework underpinning the extraction of CKM elements by incorporating next-to-leading-logarithmic (NLL) QCD corrections to the well-known electromagnetic effects within an effective field theory (EFT) approach~\cite{Gorbahn:2022rgl,Cirigliano:2023fnz,GJMvM:2026}.
A consistent inclusion of these mixed QED--QCD effects is essential to reduce the theoretical uncertainty in $V_{ud}$ and to sharpen precision tests of first-row CKM unitarity.

In our EFT framework, we consistently sum $\mathcal{O}(\alpha\, \alpha_s^2)$ log enhanced contributions to the radiative correction $\Delta^V_R$, relevant for the traditional analyses of neutron and superallowed beta decays~\cite{Hardy:2020qwl,Gorchtein:2021fce,Czarnecki:2019mwq,Gorchtein:2023naa,Ma:2023kfr,Seng:2018qru,Cirigliano:2022yyo,Marciano:2005ec,Seng:2018yzq,Seng:2022cnq,Towner:2010zz,Czarnecki:2004cw}, as well as its recently introduced~\cite{Cirigliano:2023fnz,Cirigliano:2024msg,Cirigliano:2024rfk} EFT counterpart $\Delta^V_{R,\mathrm{EFT}}$. Including these higher-order corrections reduces the residual scheme and scale dependence, order by order in perturbation theory.
We study the resulting reduction in scale dependence and systematically assess the convergence of higher-order corrections in the extraction of $V_{ud}$, thus laying the groundwork for future precision determinations.

The Heavy Baryon Chiral Perturbation Theory (\HBchiPT{}) effective Hamiltonian~\cite{Ando:2004rk,Raha:2011aa,Falkowski:2021vdg} for nuclear beta decay reads
\begin{equation}
  \label{eq:pionless-lagrangian}
  \mathcal{H}_{\slashed{\pi}} = \sqrt{2} G_F V_{ud} \bar{e} \gamma_\mu P_L\nu_{e} \bar{N}_v \left(  g_V v^\mu - 2 g_A S^\mu \right) \tau^+  N_v,
\end{equation}
where $N_v = \left( p,~n \right)^T$ is the heavy-nucleon field doublet, $v^\mu$ the nucleon velocity, $S^\mu = \left( 0, \vec\tau \right)$ the nucleon spin, $\tau^+ = \tau^1 + i \tau^2$ is defined in terms of the strong isospin matrices $\vec{\tau}$, and $P_L = (1-\gamma_5)/2$. The product $V_{ud} G_F$ of the relevant CKM matrix element times the Fermi constant extracted from muon decay stems from the weak effective theory of the Standard Model.
In this theory, the effective Hamiltonian
\begin{equation}
  \label{eq:eff-wet-hamiltonian}
  \mathcal{H}=2 \sqrt{2} G_{F} V^{*}_{ud} C(\mu) O
\end{equation}
to an excellent approximation, comprises only a single charged-current operator
$O=(\bar{d}\gamma^{\mu}P_{L}u) (\bar{\nu}_{l}\gamma_{\mu}P_{L}l)$,
while current conservation implies the matching relation $g_V = 1 + \mathcal{O}(\alpha)$ between the two effective theories.

We factorize higher-order $\alpha \, \alpha_s^n$ corrections to the vector coupling into three contributions as 
\begin{equation}
  \label{eq:gv}
  g_V(\mu_{\chi}) = C(\mu) m_{\mu_0}(\mu,\mu_{\chi}) (1 + \overline{\Box}^V_{\rm Had}(\mu_0)) \,,
\end{equation}
where $C(\mu)$ is a Wilson coefficient that comprises the short-distance contributions and is evaluated in the $\msbar$ scheme at a regularization scale $\mu$, $m_{\mu_0}$ is a matching coefficient that converts to the \HBchiPT{} at a regularization scale $\mu_{\chi}$, and $\overline{\Box}^V_{\rm Had}(\mu_0)$ denotes the long-distance corrections to the matching and $\mu_0$ is a factorization scale.
Only the isoscalar contribution—proportional to the sum of the down- and up-quark charges $Q_+ = Q_d + Q_u$, receives QCD corrections~\cite{Descotes-Genon:2005wrq}.
This, together with current conservation, allows the factorization of long- and short-distance QCD corrections.

The nonisoscalar corrections, which are proportional to $Q_- = Q_d - Q_u$, can be further factorized as $m_{\mu_0}(\mu,\mu_{\chi}) = m_{\mu_0}(\mu)[1 + 3 \alpha/(8 \pi) (1-\log \mu^2_{\chi}/\mu^2)]$, while the remaining nonperturbative isoscalar corrections $\overline{\Box}^V_{\rm Had}(\mu_0)$ are equivalent to the W-photon box $\overline{\Box}_{W,\gamma}(\mu_0)$ contribution up to $\mathcal{O}(\mu_0^2/M_W^2)$ power corrections~\cite{Cirigliano:2023fnz}.
The residual dependence on the scale $\mu$ and the weak effective theory scheme  cancels at each order in perturbation theory. 
The remaining dependence on $\mu_{\chi}$ is due to QED logarithms in the \HBchiPT\ that can be resummed using chiral perturbation theory~\cite{Cirigliano:2023fnz}.

The paper is organized as follows: We first derive the matching of the weak effective theory to the \HBchiPT{} by introducing an intermediate renormalization scheme that allows us to systematically factorize nonperturbative and perturbative QCD effects. Next, we combine these matching corrections with the required three-loop anomalous dimensions and two-loop matching coefficients of the weak effective theory. This yields scheme- and scale-independent quantities, whose convergence is analyzed in our numerical study. Finally, we examine the impact of these corrections on the extraction of $V_{ud}$ and on first-row CKM unitarity before concluding.


When matching the weak effective theory onto the hadronic theory, no hard gluons are present in the latter; all short-distance QCD corrections must therefore be completely factorized.
The remaining nonperturbative long-distance corrections, parameterized by the W-photon box contribution, involve the Nachtmann moment, which is defined via an operator product expansion.
We introduce an intermediate renormalization scheme for the weak effective theory that regularizes the short-distance dynamics, factorizes the QCD corrections, and matches directly onto the W$\gamma$-box contribution.

The renormalization of the operator $O$ in the weak effective theory is complicated by the appearance of evanescent operators in the Naive Dimensional Regularization (NDR) scheme.
Their renormalization is fixed by requiring that the renormalized infrared-finite Green’s functions vanish in the four-dimensional limit, while their form for
$d\neq4$ results in a scheme dependence.
Up to $\mathcal{O}(\alpha)$, only a single evanescent operator is required, which we choose as
\begin{equation}
  \label{eq:evanescent-operator}
  E = (\bar{d}\gamma^{\mu}\gamma^{\nu}\gamma^{\lambda}P_{L}u) (\bar{\nu}_{l}\gamma_{\mu}\gamma_{\nu}\gamma_{\lambda}P_{L}l) - 4(4 -a \epsilon) O,  
\end{equation}
where $\epsilon=2-d/2$. The parameter $a$ defines a specific scheme for the evanescent operator; traditionally, $a=1$ \footnote{our definition \eqref{eq:evanescent-operator} is equivalent to the one in Ref.~\cite{Cirigliano:2023fnz} after identifying $a_{\rm Ref.~\cite{Cirigliano:2023fnz}}\to (a-3)/2$.}.

We expand the product of the bare Wilson coefficients and operators in terms of the renormalized ones as $C_{b,i} O_{b,i} \to C_i Z_{ij} O_j Z_{\psi}$, where we denote by $Z_{\psi}$ the product of the fermion field renormalization and sum over $i,j \in\{O,E\}$ (we identify the 4 dimensional objects $C_O$ and $O_O$ with $C$ and $O$ to simplify notation).
We obtain the renormalization constant up to $\mathcal{O}(\alpha \alpha_s)$
\begin{equation}
  \label{eq:z-oopm}
  \begin{split}
    Z_{OO} = & 1 + \frac{\alpha}{4\pi \epsilon}
     \left[ Q_e Q_+ \left(\frac{3}{2}-7 \frac{\alpha_s}{4\pi}\right)
      + Q_e^2 \frac{\xi}{2} \right. \\  
      & \left. -  Q_e Q_- \left( \frac{3}{2} + \xi - \frac{\alpha_s}{4\pi} \right) +
      Q_-^2 \left( \frac{\xi}{2} - \frac{\alpha_s}{4\pi} \right)  \right].
  \end{split}
\end{equation}
Here, we retain the full dependence on the fermion charges, where $Q_e$ denotes the electron charge.
The contribution of the quark field renormalization is absorbed in the term proportional to $Q_+^2$, while the electron field renormalization is proportional to $Q_e$.
Applying the charge conservation, $Q_e = Q_-$, cancels the QED gauge-fixing parameter $\xi$, leaving only the isoscalar term proportional to $Q_e Q_+$ to receive QCD corrections.
Finally, the relevant renormalization constants involving the evanescent operator read
\begin{equation}
  \label{eq:zoe-zeo}
  Z_{OE} = \frac{\alpha}{4 \pi \epsilon} Q_e Q_+ \left[ \frac{1}{4} - \frac{7}{6} \frac{\alpha_s}{4\pi} \right]
  \text{ and }
  Z_{EO} = - 32 \frac{\alpha_s}{4 \pi} .
\end{equation}

Since only the isoscalar contributions receive $\alpha_s$ renormalization, the remaining terms can be matched directly onto $\chi$PT~\cite{Cirigliano:2023fnz}.
This will cancel the $\mathcal{O}(\alpha)$ logs that correspond to the QED interactions of the proton and electron, which are proportional to $Q_e Q_-$.
In the following, we will regularise the isoscalar UV divergences by subtracting the large-momentum contribution of the photon loop.
The operator product expansion of
\begin{equation}
  \label{eq:T-VW-0}
  T^{\mu\nu}_{0}(q^2)=\int\;{\rm d}^dx\;e^{iq\cdot x} \text{T}\left[J_W^{\mu}(x)J_{V,0}^{\nu}(0)\right],
\end{equation}
where $J^{\mu}_W = \bar{d}\gamma^{\mu} P_Lu$ is the weak charged-current and $J^{\nu}_{V,0} = Q_+/2 (\bar{u} \gamma^{\nu} u + \bar{d} \gamma^{\nu} d)$ is the isoscalar electromagnetic current, reproduces the required large-momentum behavior. 

\begin{figure}[t]
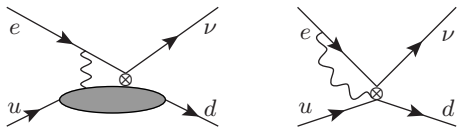

\begin{center}
  \begin{axopicture}(170,50)
  \SetColor{Black}
  \Line[arrow](0,0)(20,10)
  \Text(3,7){$u$}
  \Line[arrow](60,10)(80,0)
  \Text(77,7){$d$}
  \GOval(40,10)(5,20)(0){0.6}
  \Line[arrow](0,45)(45,20)
  \Text(3,37){$e$}
  \Line[arrow](45,20)(80,45)
  \Text(77,37){$\nu$}
  \Text(45,17.45){\tiny $\boldsymbol \otimes$}
  \Photon(30,14.5)(30,28.5){2}{2.5}
  \SetOffset(110,0)
  \SetColor{Black}
  \Line[arrow](0,0)(30,10)
  \Text(3,7){$u$}
  \Line[arrow](30,10)(60,0)
  \Text(57,7){$d$}
  \Line[arrow](0,45)(30,15)
  \Text(3,35){$e$}
  \Line[arrow](30,15)(60,45)
  \Text(57,34){$\nu$}
  \Text(30,12.45){\tiny $\boldsymbol \otimes$}
\PhotonArc(29,30.7)(20,166,267.5){2}{3.8}
\end{axopicture}
\end{center}
\caption{\label{fig:mu0} Feynman diagrams for the computation of the Green's function $\langle O \rangle_{\mu_0}$, i.e. the $\mu_0$-subtracted scheme. All diagrams are evaluated in dimensional regularization. The diagram on the right-hand side represents the OPE subtraction $\langle O \rangle_{S}$, while the diagram on the left-hand side also contributes to the $\msbar$ renormalization. The latter diagrams cancel in the scheme-matching equation.}
\end{figure}

In the NDR scheme, after shifting $q^2$ by a hard scale $\mu_0^2$ we find the OPE subtraction term up to one-loop 
\begin{equation}
  \label{eq:ndr-ope}
  \begin{split}
      T^{\mu\nu}_{S,0}(q^2) & =
    \frac{Q_+ q_{\rho}}{2(q^2-\mu^2_0)}\left\{ \left[ 1 - \frac{\alpha_s}{\pi} 1_{\epsilon} \right]o^{\mu\rho\nu} \right. \\
    + & \left.
    \left[ 1 - \frac{\alpha_s}{4\pi} \left( \frac{\mu^2}{\mu_0^2-q^2} \right)^{\epsilon} \left( \frac{28}{3} + 12 \epsilon \right) \right]
    e^{\mu\rho\nu} 
    \right\},
  \end{split}
\end{equation}
where we define the physical and evanescent structures as
\begin{equation}
  \label{eq:ope-structures}
  \begin{split}
    o^{\mu\rho\nu}&= 2i\varepsilon^{\mu\nu\rho\sigma}\bar{d}\gamma^{\sigma}P_{L}u \quad \text{and} \\
    e^{\mu\rho\nu}&=\bar{d}(\gamma^{\mu} \gamma^{\rho} \gamma^{\nu} - \gamma^{\nu} \gamma^{\rho} \gamma^{\mu})P_{L}u - (1-z_{eo})o^{\mu\rho\nu}\,,
  \end{split}
\end{equation}
respectively.
The fact that $1_{\epsilon} = \frac{7 + 9 \epsilon}{3} \big( \frac{\mu^{2\epsilon}}{\mu_0^2-q^2} \big)^{\epsilon}  - \frac{4}{3} \to 1$ in the limit $\epsilon \to 0$ recovers the result of the four dimensional OPE, while
the finite renormalization $z_{eo} = - (16/3) (\alpha_s/4 \pi)$ ensures that matrix elements of the evanescent structure vanish in the limit $d \to 4$.
Matrix elements of semileptonic processes involving the subtracted operator product $T^{\mu\nu}_{\mu_0,0} = T^{\mu\nu}_0 - T^{\mu\nu}_{S,0}$ are now finite, since the subtraction term regularizes the large-momentum component of the photon loop above the scale $\mu_0$, while current conservation ensures the finiteness of the QCD subdivergence~\footnote{We confirmed at two loop that the subtraction term exactly reproduced the weak effective theory using different IR regulators such as a photon mass or off-shell momenta.}.
Accordingly, we can write the Green's function in the subtracted scheme as $\langle O \rangle_{\mu_0}=\langle O \rangle - \langle O \rangle_S$ as a sum of the bare diagrams and the subtraction term -- see Fig.~\ref{fig:mu0} for a diagrammatic representation of the subtraction.

The matrix elements of the $\msbar$ and $\mu_0$ renormalized schemes are equal 
\begin{equation}
\label{eq:scheme-change}
Z_{O,i}^{s.c.} Z_{i,j} \langle Q_j \rangle = \langle O \rangle_{\mu_0}
\end{equation}
up to a finite change of scheme $Z_{O,i}^{s.c.}$.
It follows that all loop induced matrix elements cancel in the above matching equation except for the (perturbative) subtraction terms.
Computing these subtraction terms, we find the $\mathcal{O}(\alpha)$ scheme change of the physical operator proportional to the evanescent operator 
\begin{equation}
  \label{eq:z-sce}
  Z_{O,E}^{s.c.} =  \frac{\alpha}{4\pi} \left( \frac{3}{8} - \frac{1}{2} \log \frac{\mu_0}{\mu}  \right) Q_e Q_+,
\end{equation}
which we need to determine the scheme change 
\begin{equation}
  \begin{split}
    \label{eq:z-sce}
    Z_{O,O}^{s.c.} = 1 + &\frac{\alpha}{4 \pi}  Q_e Q_+
    \left[\frac{15-4a}{4} - 3 \log \frac{\mu_0}{\mu} \right. \\ &
    \left. 
    + \frac{\alpha_s}{4\pi} \left( \frac{28 a -87}{6} + 12 \log \frac{\mu_0}{\mu} \right)  \right]
  \end{split}
\end{equation}
up to $\mathcal{O}(\alpha \alpha_s)$ from (\ref{eq:scheme-change}).
The $\mathcal{O}(\alpha)$ terms in the scheme change are consistent with the analysis of Ref.\cite{Cirigliano:2023fnz}, which were obtained by simply subtracting the OPE, while the finite $\mathcal{O}(\alpha \, \alpha_s)$ terms are new.
The inclusion of the evanescent mixing and $\epsilon$ terms of (\ref{eq:ndr-ope}) in the evaluation $\langle O \rangle_S$ are essential to obtain a finite result with the correct scheme and scale dependence.
After combining the isoscalar matching $m_{\mu_0}(\mu) = (Z^{s.c.}_{O,O})^{-1}$ with the nonisoscalar contribution, we find that the resulting matching factor $m_{\mu_0}(\mu, \mu_{\chi})$ cancels the remaining $\log \mu_{\chi}$ dependence of \HBchiPT, thereby achieving complete factorization of the short-distance QCD corrections originating from the isoscalar contribution.

The remaining long-distance isoscalar contribution is given as an integral over the $\mu_0$-regularized operator product with external baryon and leptons evaluated in \HBchiPT.
As this integral over the photon-loop momentum is finite for all values of $d$, we can simply take the limit $d \to 4$ where then also the matrix elements involving the  evanescent structures vanish in this limit.
Only the vector-axial component of the weak hadronic current contributes to these integrals~\cite{Gorchtein:2023naa,Cirigliano:2023fnz}.
Defining $Q^2=-q^2$, the neutron mass $m_N$, $\nu = v \cdot q$ and the amplitude $T_3$ as the amplitude of the tensor decomposition of the hadronic tensor $T^{\mu\nu}_{VA,0} = i \epsilon^{\mu\nu\sigma\rho}q_{\rho}v_{\sigma} T_3/(4 m_N \nu) + \dots$ 
we find 
\begin{equation}
  \label{eq:hadronic-box}
  \begin{split}
    \overline{\Box}^V_{\rm Had}(\mu_0)&=-e^2\int \frac{i{\rm d}^4q}{(2\pi)^4} \frac{\nu^2+Q^2}{Q^4} \frac{T_3(\nu,Q^2)}{2m_N\nu} \\
    +& e^2\int \frac{i{\rm d}^4q}{(2\pi)^4}
    \frac{2}{3}\frac{\nu^2+Q^2}{Q^6+\mu_0^2Q^4}\left( 1-\frac{\alpha_s(\mu_0^2)}{\pi} \right) \,,
  \end{split}
\end{equation}
which is directly related to the $\gamma{}W$ box function.
Its form follows from the required scale cancellation and agrees with the $d=4$ derivation of Ref.~\cite{Cirigliano:2023fnz}. Our NDR derivation yields a consistent factorization of short-distance QCD effects in Eq.~(\ref{eq:z-sce}).

The first integral in Eq.~(\ref{eq:hadronic-box}) can be transformed \cite{Seng:2018yzq,Seng:2018qru} into a one-dimensional integral over $F(Q^2)=(12/Q^2) M_3^{(0)}(1,Q^2)$, where $M_3^{(0)}(1,Q^2)$ is the first Nachtmann moment, and split into a low ($Q^2<Q_0^2$) and high ($Q^2>Q_0^2$) region for the integration variable $Q^2>0$.
For sufficiently large $Q_0^2$ one enters the deep inelastic scattering (DIS) region and can combine the high $Q^2$ region with the second integral and finds
\begin{equation}
  \label{eq:eq:hadronic-box-q2}
  \overline{\Box}^V_{\rm Had}(\mu_0) =
  \Box^{V<}_{\gamma W}(Q_0^2) +
  \frac{\alpha}{8 \pi} \left[ 1 - \frac{\alpha_s(\mu_0)}{\pi}\right] \log \frac{\mu_0^2}{Q_0^2} ,
\end{equation}
where $\Box^{V<}_{\gamma W}(\mu_0)$ denotes the contribution of the elastic and inelastic low $Q^2$ region of the first integral that has been determined in the literature either using a dispersive approach, see Ref.~\cite{Seng:2018yzq,Seng:2018qru,Czarnecki:2019mwq,Seng:2020wjq,Hayen:2020cxh,Shiells:2020fqp,Gorchtein:2023naa} for details, or using lattice QCD~\cite{Ma:2023kfr}.
The form follows from the required scale cancellation~\cite{Cirigliano:2023fnz}.
In the NDR approach, additional higher-order corrections can be included in the terms within square brackets by combining the perturbative DIS calculation~\cite{Larin:1990zw,Larin:1991tj,Baikov:2010je,Czarnecki:2019mwq} with higher-order corrections to the OPE of Eq.~(\ref{eq:ndr-ope}).
These higher-order contributions are neglected, as they lie beyond the scope of our NLL analysis.


The dependence of the Wilson coefficient on $\mu$ is governed by the renormalization-group equation $\mu \frac{d}{d \mu}C^{(f)}(\mu) = \gamma^{(f)} C^{(f)}(\mu)$ which depends on the number of active flavors $f$.
In this work, we expand the anomalous dimension (see Refs.~\cite{Cirigliano:2023fnz} and \cite{GJMvM:2026} for details on the $\mathcal{O}(\alpha^2)$ and $\mathcal{O}(\alpha \, \alpha_s^2)$ calculations, respectively)
\begin{equation}
  \label{eq:gamma-expansion}
  \begin{split}
  \gamma^{(f)} =&
    -\frac{\alpha(\mu)}{\pi} + \left[\frac{2a-4}{9} \overline{Q^{2}_f} \right]\frac{\alpha^2(\mu)}{4 \pi^2} + \frac{\alpha \alpha_s(\mu)}{4\pi^2} + \\ &
    \left[ a \left(\frac{308}{9}-\frac{56 n_f}{27}\right)+\frac{10 n_f}{9} -33 \right]\frac{\alpha \alpha_s^2(\mu)}{(4\pi)^3}
  \end{split}
\end{equation}
up to ${\cal O}(\alpha^2)$ and ${\cal O}(\alpha\alpha_{s}^{2})$ for pure QED and mixed QED$\times$QCD corrections, respectively. Both $\alpha_s(\mu) = \alpha_s^{(f)}(\mu)$ and $\alpha(\mu) = \alpha^{(f)}(\mu)$ are defined in a theory with $f$ active quark flavors.
Here $\overline{Q^{2}_f} = \sum_f Q_f^2$ denotes a sum over active fermions with charge $Q_f$, $n_f$ is the number of active quark flavors.
We take into account the $\tau$ threshold in the running of $\alpha$, but decouple the $\tau$ lepton and charm quark simultaneously for the Wilson coefficient.

Since we do not consider $\mathcal{O}(\alpha^2 \alpha_s)$ corrections, we can neglect the scale dependence of the electromagnetic coupling constant in the mixed QED$\times$QCD contributions to the anomalous dimension.
Within this approximation, we find the following solution
\begin{equation}
\label{eq:U-definition}
C^{(f)}(\mu)= \je{f}(\mu) \ue{f}(\mu) \ue{f}^{-1} (\mu_0) \je{f}^{-1}(\mu_0) C^{(f)}(\mu_0)\,,
\end{equation}
where the leading-order evolution kernel
\begin{equation}
  \label{eq:u-defintion}
  \ue{f}(\mu) =
  \left( \frac{\alpha^{(f)}(\mu)}{\alpha(M_Z)} \right)^{-\frac{3}{2 \overline{Q^2_f}}}
  \left( \frac{\alpha^{(f)}_s(\mu)}{\alpha_s(M_Z)} \right)^{-\frac{\alpha}{4\pi}\frac{6}{33 - 2 n_f}}
\end{equation}
resums terms of order $\alpha^n \log^n$ and $\alpha \alpha_s^n \log^n$.
Here we normalized the running coupling constants to their experimental input values at $M_Z$ to reduce the dependence on these parameters.
The correction to the evolution kernel is parameterized by $J$, which equals 1 plus $\mathcal{O}(\alpha)$ and $\mathcal{O}(\alpha \alpha_s)$ corrections that represent the scheme-dependent NLL QED and QCD corrections.

The final physical result must, of course, be independent of the choice of renormalization scheme
and we define regularization-independent quantities that allow us to check this scheme dependence:
\begin{equation}
  \label{eq:define-hat}
  \begin{split}
  \hat{C} &= \ue{5}^{-1}(\mu) \je{5}^{-1}(\mu) C^{(5)}(\mu) \,, \\
  \hat{M}_{f} &= \ue{f-1}^{-1}(\mu) \je{f-1}^{-1}(\mu) \je{f}(\mu) \ue{f}(\mu) \,, \\
  \hat{M}_{\mu_{\chi}} &= \left[ 1 + \overline{\Box}^V_{\rm Had}(\mu_0) \right] m_{\mu_0}(\mu,\mu_{\chi}) \je{3}(\mu) \ue{3}(\mu) \,. \\
  \end{split}
\end{equation}
Here we introduce the flavor decoupling operator $\hat{M}_{f}$ and included the nonperturbative parameter $\overline{\Box}^V_{\rm Had}$ in the matching to the \HBchiPT.
These quantities have the additional benefit that their dependence on the regularization scale $\mu$ cancels order-by-order in perturbation theory, so that we can use their residual scale dependence as a measure of unknown higher order corrections.

Using these quantities, the renormalization-group-improved expression for the vector coupling reads
\begin{equation}
  \label{eq:rge-improved-amp}
  g_V(\mu_{\chi}) = \hat{M}_{\mu_{\chi}} \hat{M}_4 \hat{M}_5 \hat{C}^{(5)}.
\end{equation}

Defining $l^i_j = \log(m_i^2/m_j^2)$, where $m_{\mu}$ and $m_{\chi}$
denote the scale of dimensional regularization $\mu$ in the low energy
effective theory and in \HBchiPT\ $\mu_{\chi}$, respectively 
we find, using results from a two-loop electroweak matching calculation~\cite{GJMvM:2026}, for the high scale Wilson coefficient evaluated at $\mu \sim M_W$
\begin{equation}
  \label{eq:c-hat}
  \begin{split}
    \hat{C} =& u^{-1}_{5}(\mu)
    \left( 1 -
\frac{\alpha^{(5)}(\mu)}{4\pi}
  \left[
    2 \lmz +
    \frac{2261}{600}
  \right]
    \right. + \\ &
    \left.
    \frac{\alpha \alpha_s^{(5)}(\mu)}{(4\pi)^2}  
    \left[
      \frac{1972}{529} - \frac{4}{\swsq} + 2 \lmz -
      \frac{2 \cwsq}{\swft} \lwz
    \right]  \right)\,,
  \end{split}
\end{equation}
and the matching to \HBchiPT\ evaluated at $\mu \sim \mu_0$.
\begin{eqnarray}
  \label{eq:m-hat-nlle}
  \begin{split}
    \hat{M}_{\mu_{\chi}} = u_{3}&(\mu) \left( 1 +
    \frac{\alpha^{(3)}(\mu)}{4\pi} 
  \left[ \frac{17}{6} -
    \frac{3}{2} \lchiO  + 2 \lmO
     \right]  \right. \\ & + \overline{\Box}^V_{\rm Had}(\mu_0) +
   \left. \frac{\alpha \alpha_s^{(3)}(\mu)}{(4\pi)^2}
       \left[-\frac{130}{81} - 2 \lmO \right] \right) .
  \end{split}
\end{eqnarray}
The decoupling of the bottom quark
\begin{equation}
  \label{eq:mdec5}
  \hat{M}_{5} = \frac{u_5(\mu)}{u_{4}(\mu)}\left[1-\frac{28496 \alpha  \alpha^{(4)}_s(\mu )}{330625 (4\pi)^2}-\frac{3093 \alpha^{(4)}(\mu )}{72200 (4\pi)} \right]
\end{equation}
and of the tau lepton and charm quark
\begin{equation}
  \label{eq:mdec4}
  \hat{M}_{4} = \frac{u_{4}(\mu)}{u_{3}(\mu)} \left[1 -\frac{1856 \alpha  \alpha^{(3)}_s(\mu )}{50625 (4\pi)^2}-\frac{393 \alpha^{(3)}(\mu )}{1444 (4\pi)} \right]
\end{equation}
are evaluated at the scales $\mu \sim m_b$ and $\mu \sim m_c$, respectively.
So far, only the one-loop $\mathcal{O}(\alpha)$ Wilson coefficient was known~\cite{Gambino:2001au,Brod:2008ss}. In our work, we derive the two-loop $\mathcal{O}(\alpha\alpha_{s})$ contribution. 


\begin{figure}[htb]
  \includegraphics[width=\columnwidth]{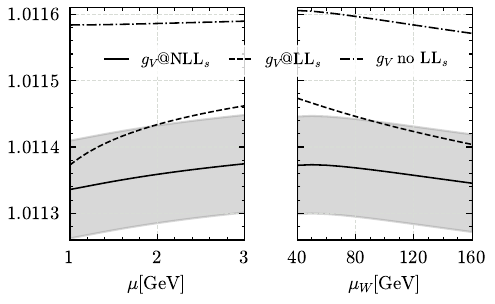}
  \caption{
    Both panels show the residual scale dependence on $\mu$ (left panel) and $\mu_W$ (right panel) for $g_V(m_n)$ at LL and NLL in QCD.
    The impact of the Lattice uncertainties in $\Box^{V<}_{\gamma W}(Q_0^2)$ are shown for the NLO curve as the grey region.
    Leading Log QCD corrections are absent in the no LL$_s$ curve.}
  \label{fig:gvmu2}
\end{figure}
For the numerical evaluation of $g_V(m_n)$ in Eq.~(\ref{eq:rge-improved-amp}), we determine the individual contributions including all NLL electromagnetic corrections and  both LL and NLL strong corrections to the LL QED contributions, denoted by $\LLs$ and $\NLLs$, respectively.
Using PDG input we find
\begin{equation}
  \label{eq:3}
  \begin{split}
  (\hat{M}_{\mu_{\chi}}-1) \cdot 10^6 &\overset{\LLs}{=}  \,20653^{+37}_{-53} \overset{\NLLs}{\to} \, 20614^{+20+74}_{-19-74} \,, \\
  (\hat{M}_4 - 1) \cdot 10^6        &\overset{\LLs}{=} \;-6266^{+3}_{-3}    \overset{\NLLs}{\to} \; -6267^{+3}_{-3}   \,, \\
  (\hat{M}_5 -1) \cdot 10^6         &\overset{\LLs}{=} \;\;\;-432^{+1}_{-1} \overset{\NLLs}{\to} \;\;\; -433^{+1}_{-1}\,, \\
  (\hat{C} - 1) \cdot 10^6     &\overset{\LLs}{=} - 2361^{+47}_{-22}   \overset{\NLLs}{\to} \!- 2391^{+13}_{-13} \,, \\
  (g_{V} - 1) \cdot 10^6     &\overset{\LLs}{=} \; 11426^{+60}_{-57}   \overset{\NLLs}{\to} \, 11355^{+24+74}_{-23-74} \,, 
  \end{split}
\end{equation}

where the first uncertainty denotes the uncertainties due to unaccounted higher-order corrections, which we estimate through a scale variation over the ranges $\mu, \mu_c \in [1,3]$GeV, $\mu_b \in [2.5,10]$GeV, and $\mu_W \in [40,160]$GeV.
Central values correspond to $\mu = \mu_c = 1.8$, $\mu_b=5$, and $\mu_W=115$, which leads to approximately symmetric uncertainties from higher-order QCD corrections.
We also set $\mu_0^2 = Q_0^2 = 2 \mathrm{GeV}^2$, but note that both the $\mu_0$ and $Q_0$ dependence would cancel in $\hat{M}_{\mu_{\chi}}$.
The second uncertainty arises from the lattice calculation~\cite{Ma:2023kfr} of $\Box^{V<}_{\gamma W}(Q_0^2) = 1.490(73)\cdot 10^{-3}$ and is only shown in our $\NLLs$ evaluations.
The NLL corrections are comparable in size to the current lattice uncertainty, underscoring the relevance of higher-order effects.

\begin{figure}[htb]
  \includegraphics[width=\columnwidth]{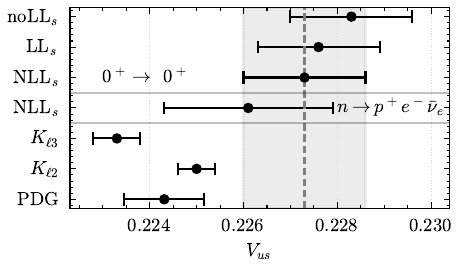}
  \caption{Determination of $V_{us}$ from CKM unitarity using the
    $\LLs$ and $\NLLs$ determinations of $V_{ud}$ for superallowed nuclear beta decays ($0^+ \to 0^+$), compared with the free neutron decay at $\NLLs$ and with $K_{\ell 2}$ and $K_{\ell 3}$ decays, including their average. The NLL result
    reduces the theoretical uncertainty of $\Delta^V_R$ and restores
    consistency with first-row CKM unitarity. Leading Log QCD
    corrections are absent in the no $\LLs$ interval.}
  \label{fig:vus-comparision}
\end{figure}

A full effective field theory analysis for neutron and superallowed nuclear beta decays would involve the evaluation of QED corrections to $g_V$ in \HBchiPT, which would involve renormalization-group effects \cite{Cirigliano:2023fnz,Borah:2024ghn} and the evaluation of soft and hard QED corrections~\cite{Hill:2023acw,Hill:2023bfh,Plestid:2024eib,Cirigliano:2024msg,Cirigliano:2024rfk,VanderGriend:2025mdc,Cao:2025lrw}. In order to extract $|V_{ud}|^2$, we combine our results with the global analysis of superallowed nuclear beta decays~\cite{Hardy:2020qwl} and follow the definition \cite{Cirigliano:2023fnz} of transition-independent part of the radiative correction and find for the traditional value~\footnote{Using the nonperturbative input $\Box^{V<}_{\gamma W}(Q_0^2) = 1.560(110)(10)(48)\cdot 10^{-3}$ from Ref.~\cite{Cirigliano:2023fnz}, which sums the Regge, resonance, and elastic contribution with errors in brackets, we find $(g_{V} - 1)_{\LLs} \cdot 10^6 = 11478^{+60}_{-57}$, corresponding to a central value of $\Delta^V_R = 2.461\%$, in agreement with their value within perturbative uncertainties.}:
\begin{equation}
\label{eq:deltavrtrad}
\Delta_R^V =
g_V^2(m_n) \left( 1 + \frac{5 \alpha(m_n)}{8\pi}\right)-1 = 2.436(16)\%\,.
\end{equation}
Using ${|V_{ud}|^{2,0^+\to 0^+} = 0.97142(58)/(1+\Delta_R^V)}$, where the numerical factor is computed from the ratio $K/(2 G_F \overline{{\cal F}t})$ using PDG~\cite{ParticleDataGroup:2024cfk} input, we find $|V_{ud}|^{0^+\to 0^+}_{\NLLs} = 0.97382(30)$.
When extracting $|V_{ud}|^2$ from the neutron lifetime, we combine our results with the radiative corrections specific to the neutron decay, $\Delta_R$, computed in Refs.~\cite{Cirigliano:2023fnz,VanderGriend:2025mdc} including two-loop input to the long-distance corrections from~\cite{VanderGriend:2025mdc}, the lattice determination~\cite{Ma:2023kfr} of the non perturbative input, and obtain $|V_{ud}|^{n}_{\NLLs}=0.97410(41)$.
Assuming CKM unitarity, we translate these extractions of $|V_{ud}|$ into a determination of $|V_{us}|$, thereby testing the internal consistency of the CKM framework. The resulting values are shown in Fig.~\ref{fig:vus-comparision} and show how the computed higher-order corrections reduce the tension of the first-row CKM unitarity. 

In summary, we have presented the first next-to-leading-logarithmic QCD analysis of electromagnetic corrections to the semileptonic weak Hamiltonian, including the mixed 
$\mathcal{O}(\alpha\,\alpha_s^2)$ corrections to the vector coupling $g_V$. 
The resulting framework consistently factorizes short- and long-distance QCD effects and reduces the theoretical uncertainty in the radiative correction $\Delta^V_R$, thereby improving the consistency of first-row CKM unitarity tests.

\section{Acknowledgment}
The work of MG is partially supported by the UK Science and Technology Facilities Council
grant ST/X000699/1. The work of FM is supported by the Deutsche Forschungsgemeinschaft (DFG,German Research Foundation) under grant 396021762 --- TRR 257 ``Particle Physics Phenomenology after the Higgs Discovery''. The work of SJ is supported in part by the UK Science and Technology Facilities Council grant ST/X000796/1.
The Feynman diagrams were drawn with the help of Axodraw~\cite{Vermaseren:1994je}. The authors would like to thank Vincenzo Cirigliano and Martin Hoferichter for comments on the draft. The authors would also like to thank Ulrich Nierste and Paul Rakow for useful discussions. 

\bibliography{bibliography}

\begin{thebibliography}{58}%
\makeatletter
\providecommand \@ifxundefined [1]{%
 \@ifx{#1\undefined}
}%
\providecommand \@ifnum [1]{%
 \ifnum #1\expandafter \@firstoftwo
 \else \expandafter \@secondoftwo
 \fi
}%
\providecommand \@ifx [1]{%
 \ifx #1\expandafter \@firstoftwo
 \else \expandafter \@secondoftwo
 \fi
}%
\providecommand \natexlab [1]{#1}%
\providecommand \enquote  [1]{``#1''}%
\providecommand \bibnamefont  [1]{#1}%
\providecommand \bibfnamefont [1]{#1}%
\providecommand \citenamefont [1]{#1}%
\providecommand \href@noop [0]{\@secondoftwo}%
\providecommand \href [0]{\begingroup \@sanitize@url \@href}%
\providecommand \@href[1]{\@@startlink{#1}\@@href}%
\providecommand \@@href[1]{\endgroup#1\@@endlink}%
\providecommand \@sanitize@url [0]{\catcode `\\12\catcode `\$12\catcode
  `\&12\catcode `\#12\catcode `\^12\catcode `\_12\catcode `\%12\relax}%
\providecommand \@@startlink[1]{}%
\providecommand \@@endlink[0]{}%
\providecommand \url  [0]{\begingroup\@sanitize@url \@url }%
\providecommand \@url [1]{\endgroup\@href {#1}{\urlprefix }}%
\providecommand \urlprefix  [0]{URL }%
\providecommand \Eprint [0]{\href }%
\providecommand \doibase [0]{https://doi.org/}%
\providecommand \selectlanguage [0]{\@gobble}%
\providecommand \bibinfo  [0]{\@secondoftwo}%
\providecommand \bibfield  [0]{\@secondoftwo}%
\providecommand \translation [1]{[#1]}%
\providecommand \BibitemOpen [0]{}%
\providecommand \bibitemStop [0]{}%
\providecommand \bibitemNoStop [0]{.\EOS\space}%
\providecommand \EOS [0]{\spacefactor3000\relax}%
\providecommand \BibitemShut  [1]{\csname bibitem#1\endcsname}%
\let\auto@bib@innerbib\@empty
\bibitem [{\citenamefont {Navas}\ \emph {et~al.}(2024)\citenamefont {Navas}
  \emph {et~al.}}]{ParticleDataGroup:2024cfk}%
  \BibitemOpen
  \bibfield  {author} {\bibinfo {author} {\bibfnamefont {S.}~\bibnamefont
  {Navas}} \emph {et~al.} (\bibinfo {collaboration} {Particle Data Group}),\
  }\bibfield  {title} {\bibinfo {title} {{Review of particle physics}},\ }\href
  {https://doi.org/10.1103/PhysRevD.110.030001} {\bibfield  {journal} {\bibinfo
   {journal} {Phys. Rev. D}\ }\textbf {\bibinfo {volume} {110}},\ \bibinfo
  {pages} {030001} (\bibinfo {year} {2024})}\BibitemShut {NoStop}%
\bibitem [{\citenamefont {Belfatto}\ \emph {et~al.}(2020)\citenamefont
  {Belfatto}, \citenamefont {Beradze},\ and\ \citenamefont
  {Berezhiani}}]{Belfatto:2019swo}%
  \BibitemOpen
  \bibfield  {author} {\bibinfo {author} {\bibfnamefont {B.}~\bibnamefont
  {Belfatto}}, \bibinfo {author} {\bibfnamefont {R.}~\bibnamefont {Beradze}},\
  and\ \bibinfo {author} {\bibfnamefont {Z.}~\bibnamefont {Berezhiani}},\
  }\bibfield  {title} {\bibinfo {title} {{The CKM unitarity problem: A trace of
  new physics at the TeV scale?}},\ }\href
  {https://doi.org/10.1140/epjc/s10052-020-7691-6} {\bibfield  {journal}
  {\bibinfo  {journal} {Eur. Phys. J. C}\ }\textbf {\bibinfo {volume} {80}},\
  \bibinfo {pages} {149} (\bibinfo {year} {2020})},\ \Eprint
  {https://arxiv.org/abs/1906.02714} {arXiv:1906.02714 [hep-ph]} \BibitemShut
  {NoStop}%
\bibitem [{\citenamefont {Coutinho}\ \emph {et~al.}(2020)\citenamefont
  {Coutinho}, \citenamefont {Crivellin},\ and\ \citenamefont
  {Manzari}}]{Coutinho:2019aiy}%
  \BibitemOpen
  \bibfield  {author} {\bibinfo {author} {\bibfnamefont {A.~M.}\ \bibnamefont
  {Coutinho}}, \bibinfo {author} {\bibfnamefont {A.}~\bibnamefont
  {Crivellin}},\ and\ \bibinfo {author} {\bibfnamefont {C.~A.}\ \bibnamefont
  {Manzari}},\ }\bibfield  {title} {\bibinfo {title} {{Global Fit to Modified
  Neutrino Couplings and the Cabibbo-Angle Anomaly}},\ }\href
  {https://doi.org/10.1103/PhysRevLett.125.071802} {\bibfield  {journal}
  {\bibinfo  {journal} {Phys. Rev. Lett.}\ }\textbf {\bibinfo {volume} {125}},\
  \bibinfo {pages} {071802} (\bibinfo {year} {2020})},\ \Eprint
  {https://arxiv.org/abs/1912.08823} {arXiv:1912.08823 [hep-ph]} \BibitemShut
  {NoStop}%
\bibitem [{\citenamefont {Belfatto}\ and\ \citenamefont
  {Berezhiani}(2021)}]{Belfatto:2021jhf}%
  \BibitemOpen
  \bibfield  {author} {\bibinfo {author} {\bibfnamefont {B.}~\bibnamefont
  {Belfatto}}\ and\ \bibinfo {author} {\bibfnamefont {Z.}~\bibnamefont
  {Berezhiani}},\ }\bibfield  {title} {\bibinfo {title} {{Are the CKM anomalies
  induced by vector-like quarks? Limits from flavor changing and Standard Model
  precision tests}},\ }\href {https://doi.org/10.1007/JHEP10(2021)079}
  {\bibfield  {journal} {\bibinfo  {journal} {JHEP}\ }\textbf {\bibinfo
  {volume} {10}},\ \bibinfo {pages} {079}},\ \Eprint
  {https://arxiv.org/abs/2103.05549} {arXiv:2103.05549 [hep-ph]} \BibitemShut
  {NoStop}%
\bibitem [{\citenamefont {Cheung}\ \emph {et~al.}(2020)\citenamefont {Cheung},
  \citenamefont {Keung}, \citenamefont {Lu},\ and\ \citenamefont
  {Tseng}}]{Cheung:2020vqm}%
  \BibitemOpen
  \bibfield  {author} {\bibinfo {author} {\bibfnamefont {K.}~\bibnamefont
  {Cheung}}, \bibinfo {author} {\bibfnamefont {W.-Y.}\ \bibnamefont {Keung}},
  \bibinfo {author} {\bibfnamefont {C.-T.}\ \bibnamefont {Lu}},\ and\ \bibinfo
  {author} {\bibfnamefont {P.-Y.}\ \bibnamefont {Tseng}},\ }\bibfield  {title}
  {\bibinfo {title} {{Vector-like Quark Interpretation for the CKM Unitarity
  Violation, Excess in Higgs Signal Strength, and Bottom Quark Forward-Backward
  Asymmetry}},\ }\href {https://doi.org/10.1007/JHEP05(2020)117} {\bibfield
  {journal} {\bibinfo  {journal} {JHEP}\ }\textbf {\bibinfo {volume} {05}},\
  \bibinfo {pages} {117}},\ \Eprint {https://arxiv.org/abs/2001.02853}
  {arXiv:2001.02853 [hep-ph]} \BibitemShut {NoStop}%
\bibitem [{\citenamefont {Branco}\ \emph {et~al.}(2021)\citenamefont {Branco},
  \citenamefont {Penedo}, \citenamefont {Pereira}, \citenamefont {Rebelo},\
  and\ \citenamefont {Silva-Marcos}}]{Branco:2021vhs}%
  \BibitemOpen
  \bibfield  {author} {\bibinfo {author} {\bibfnamefont {G.~C.}\ \bibnamefont
  {Branco}}, \bibinfo {author} {\bibfnamefont {J.~T.}\ \bibnamefont {Penedo}},
  \bibinfo {author} {\bibfnamefont {P.~M.~F.}\ \bibnamefont {Pereira}},
  \bibinfo {author} {\bibfnamefont {M.~N.}\ \bibnamefont {Rebelo}},\ and\
  \bibinfo {author} {\bibfnamefont {J.~I.}\ \bibnamefont {Silva-Marcos}},\
  }\bibfield  {title} {\bibinfo {title} {{Addressing the CKM unitarity problem
  with a vector-like up quark}},\ }\href
  {https://doi.org/10.1007/JHEP07(2021)099} {\bibfield  {journal} {\bibinfo
  {journal} {JHEP}\ }\textbf {\bibinfo {volume} {07}},\ \bibinfo {pages}
  {099}},\ \Eprint {https://arxiv.org/abs/2103.13409} {arXiv:2103.13409
  [hep-ph]} \BibitemShut {NoStop}%
\bibitem [{\citenamefont {Crivellin}\ \emph {et~al.}(2020)\citenamefont
  {Crivellin}, \citenamefont {Kirk}, \citenamefont {Manzari},\ and\
  \citenamefont {Montull}}]{Crivellin:2020ebi}%
  \BibitemOpen
  \bibfield  {author} {\bibinfo {author} {\bibfnamefont {A.}~\bibnamefont
  {Crivellin}}, \bibinfo {author} {\bibfnamefont {F.}~\bibnamefont {Kirk}},
  \bibinfo {author} {\bibfnamefont {C.~A.}\ \bibnamefont {Manzari}},\ and\
  \bibinfo {author} {\bibfnamefont {M.}~\bibnamefont {Montull}},\ }\bibfield
  {title} {\bibinfo {title} {{Global Electroweak Fit and Vector-Like Leptons in
  Light of the Cabibbo Angle Anomaly}},\ }\href
  {https://doi.org/10.1007/JHEP12(2020)166} {\bibfield  {journal} {\bibinfo
  {journal} {JHEP}\ }\textbf {\bibinfo {volume} {12}},\ \bibinfo {pages}
  {166}},\ \Eprint {https://arxiv.org/abs/2008.01113} {arXiv:2008.01113
  [hep-ph]} \BibitemShut {NoStop}%
\bibitem [{\citenamefont {Kirk}(2021)}]{Kirk:2020wdk}%
  \BibitemOpen
  \bibfield  {author} {\bibinfo {author} {\bibfnamefont {M.}~\bibnamefont
  {Kirk}},\ }\bibfield  {title} {\bibinfo {title} {{Cabibbo anomaly versus
  electroweak precision tests: An exploration of extensions of the Standard
  Model}},\ }\href {https://doi.org/10.1103/PhysRevD.103.035004} {\bibfield
  {journal} {\bibinfo  {journal} {Phys. Rev. D}\ }\textbf {\bibinfo {volume}
  {103}},\ \bibinfo {pages} {035004} (\bibinfo {year} {2021})},\ \Eprint
  {https://arxiv.org/abs/2008.03261} {arXiv:2008.03261 [hep-ph]} \BibitemShut
  {NoStop}%
\bibitem [{\citenamefont {Crivellin}\ \emph
  {et~al.}(2021{\natexlab{a}})\citenamefont {Crivellin}, \citenamefont
  {Hoferichter},\ and\ \citenamefont {Manzari}}]{Crivellin:2021njn}%
  \BibitemOpen
  \bibfield  {author} {\bibinfo {author} {\bibfnamefont {A.}~\bibnamefont
  {Crivellin}}, \bibinfo {author} {\bibfnamefont {M.}~\bibnamefont
  {Hoferichter}},\ and\ \bibinfo {author} {\bibfnamefont {C.~A.}\ \bibnamefont
  {Manzari}},\ }\bibfield  {title} {\bibinfo {title} {{Fermi Constant from Muon
  Decay Versus Electroweak Fits and Cabibbo-Kobayashi-Maskawa Unitarity}},\
  }\href {https://doi.org/10.1103/PhysRevLett.127.071801} {\bibfield  {journal}
  {\bibinfo  {journal} {Phys. Rev. Lett.}\ }\textbf {\bibinfo {volume} {127}},\
  \bibinfo {pages} {071801} (\bibinfo {year} {2021}{\natexlab{a}})},\ \Eprint
  {https://arxiv.org/abs/2102.02825} {arXiv:2102.02825 [hep-ph]} \BibitemShut
  {NoStop}%
\bibitem [{\citenamefont {Crivellin}\ and\ \citenamefont
  {Hoferichter}(2020)}]{Crivellin:2020lzu}%
  \BibitemOpen
  \bibfield  {author} {\bibinfo {author} {\bibfnamefont {A.}~\bibnamefont
  {Crivellin}}\ and\ \bibinfo {author} {\bibfnamefont {M.}~\bibnamefont
  {Hoferichter}},\ }\bibfield  {title} {\bibinfo {title} {{\ensuremath{\beta}
  Decays as Sensitive Probes of Lepton Flavor Universality}},\ }\href
  {https://doi.org/10.1103/PhysRevLett.125.111801} {\bibfield  {journal}
  {\bibinfo  {journal} {Phys. Rev. Lett.}\ }\textbf {\bibinfo {volume} {125}},\
  \bibinfo {pages} {111801} (\bibinfo {year} {2020})},\ \Eprint
  {https://arxiv.org/abs/hep-ph/2002.07184} {arXiv:hep-ph/2002.07184 [hep-ph]}
  \BibitemShut {NoStop}%
\bibitem [{\citenamefont {Capdevila}\ \emph {et~al.}(2021)\citenamefont
  {Capdevila}, \citenamefont {Crivellin}, \citenamefont {Manzari},\ and\
  \citenamefont {Montull}}]{Capdevila:2020rrl}%
  \BibitemOpen
  \bibfield  {author} {\bibinfo {author} {\bibfnamefont {B.}~\bibnamefont
  {Capdevila}}, \bibinfo {author} {\bibfnamefont {A.}~\bibnamefont
  {Crivellin}}, \bibinfo {author} {\bibfnamefont {C.~A.}\ \bibnamefont
  {Manzari}},\ and\ \bibinfo {author} {\bibfnamefont {M.}~\bibnamefont
  {Montull}},\ }\bibfield  {title} {\bibinfo {title} {{Explaining $b\to
  s\ell^+\ell^-$ and the Cabibbo angle anomaly with a vector triplet}},\ }\href
  {https://doi.org/10.1103/PhysRevD.103.015032} {\bibfield  {journal} {\bibinfo
   {journal} {Phys. Rev. D}\ }\textbf {\bibinfo {volume} {103}},\ \bibinfo
  {pages} {015032} (\bibinfo {year} {2021})},\ \Eprint
  {https://arxiv.org/abs/2005.13542} {arXiv:2005.13542 [hep-ph]} \BibitemShut
  {NoStop}%
\bibitem [{\citenamefont {Crivellin}\ and\ \citenamefont
  {Hoferichter}(2021)}]{Crivellin:2021sff}%
  \BibitemOpen
  \bibfield  {author} {\bibinfo {author} {\bibfnamefont {A.}~\bibnamefont
  {Crivellin}}\ and\ \bibinfo {author} {\bibfnamefont {M.}~\bibnamefont
  {Hoferichter}},\ }\bibfield  {title} {\bibinfo {title} {{Hints of lepton
  flavor universality violations}},\ }\href
  {https://doi.org/10.1126/science.abk2450} {\bibfield  {journal} {\bibinfo
  {journal} {Science}\ }\textbf {\bibinfo {volume} {374}},\ \bibinfo {pages}
  {1051} (\bibinfo {year} {2021})},\ \Eprint {https://arxiv.org/abs/2111.12739}
  {arXiv:2111.12739 [hep-ph]} \BibitemShut {NoStop}%
\bibitem [{\citenamefont {Crivellin}\ \emph
  {et~al.}(2021{\natexlab{b}})\citenamefont {Crivellin}, \citenamefont
  {Manzari}, \citenamefont {Alguero},\ and\ \citenamefont
  {Matias}}]{Crivellin:2020oup}%
  \BibitemOpen
  \bibfield  {author} {\bibinfo {author} {\bibfnamefont {A.}~\bibnamefont
  {Crivellin}}, \bibinfo {author} {\bibfnamefont {C.~A.}\ \bibnamefont
  {Manzari}}, \bibinfo {author} {\bibfnamefont {M.}~\bibnamefont {Alguero}},\
  and\ \bibinfo {author} {\bibfnamefont {J.}~\bibnamefont {Matias}},\
  }\bibfield  {title} {\bibinfo {title} {{Combined Explanation of the $Z\to
  b\bar{b}$ Forward-Backward Asymmetry, the Cabibbo Angle Anomaly, and
  \ensuremath{\tau}\textrightarrow{}\ensuremath{\mu}\ensuremath{\nu}\ensuremath{\nu}
  and b\textrightarrow{}s\ensuremath{\ell^+}\ensuremath{\ell^-} Data}},\ }\href
  {https://doi.org/10.1103/PhysRevLett.127.011801} {\bibfield  {journal}
  {\bibinfo  {journal} {Phys. Rev. Lett.}\ }\textbf {\bibinfo {volume} {127}},\
  \bibinfo {pages} {011801} (\bibinfo {year} {2021}{\natexlab{b}})},\ \Eprint
  {https://arxiv.org/abs/2010.14504} {arXiv:2010.14504 [hep-ph]} \BibitemShut
  {NoStop}%
\bibitem [{\citenamefont {Marzocca}\ and\ \citenamefont
  {Trifinopoulos}(2021)}]{Marzocca:2021azj}%
  \BibitemOpen
  \bibfield  {author} {\bibinfo {author} {\bibfnamefont {D.}~\bibnamefont
  {Marzocca}}\ and\ \bibinfo {author} {\bibfnamefont {S.}~\bibnamefont
  {Trifinopoulos}},\ }\bibfield  {title} {\bibinfo {title} {{Minimal
  Explanation of Flavor Anomalies: B-Meson Decays, Muon Magnetic Moment, and
  the Cabibbo Angle}},\ }\href {https://doi.org/10.1103/PhysRevLett.127.061803}
  {\bibfield  {journal} {\bibinfo  {journal} {Phys. Rev. Lett.}\ }\textbf
  {\bibinfo {volume} {127}},\ \bibinfo {pages} {061803} (\bibinfo {year}
  {2021})},\ \Eprint {https://arxiv.org/abs/2104.05730} {arXiv:2104.05730
  [hep-ph]} \BibitemShut {NoStop}%
\bibitem [{\citenamefont {Cirigliano}\ \emph {et~al.}(2010)\citenamefont
  {Cirigliano}, \citenamefont {Jenkins},\ and\ \citenamefont
  {Gonzalez-Alonso}}]{Cirigliano:2009wk}%
  \BibitemOpen
  \bibfield  {author} {\bibinfo {author} {\bibfnamefont {V.}~\bibnamefont
  {Cirigliano}}, \bibinfo {author} {\bibfnamefont {J.}~\bibnamefont
  {Jenkins}},\ and\ \bibinfo {author} {\bibfnamefont {M.}~\bibnamefont
  {Gonzalez-Alonso}},\ }\bibfield  {title} {\bibinfo {title} {{Semileptonic
  decays of light quarks beyond the Standard Model}},\ }\href
  {https://doi.org/10.1016/j.nuclphysb.2009.12.020} {\bibfield  {journal}
  {\bibinfo  {journal} {Nucl. Phys. B}\ }\textbf {\bibinfo {volume} {830}},\
  \bibinfo {pages} {95} (\bibinfo {year} {2010})},\ \Eprint
  {https://arxiv.org/abs/hep-ph/0908.1754} {arXiv:hep-ph/0908.1754 [hep-ph]}
  \BibitemShut {NoStop}%
\bibitem [{\citenamefont {Alok}\ \emph {et~al.}(2023)\citenamefont {Alok},
  \citenamefont {Dighe}, \citenamefont {Gangal},\ and\ \citenamefont
  {Kumar}}]{Alok:2021ydy}%
  \BibitemOpen
  \bibfield  {author} {\bibinfo {author} {\bibfnamefont {A.~K.}\ \bibnamefont
  {Alok}}, \bibinfo {author} {\bibfnamefont {A.}~\bibnamefont {Dighe}},
  \bibinfo {author} {\bibfnamefont {S.}~\bibnamefont {Gangal}},\ and\ \bibinfo
  {author} {\bibfnamefont {J.}~\bibnamefont {Kumar}},\ }\bibfield  {title}
  {\bibinfo {title} {{Leptonic operators for the Cabbibo angle anomaly with
  SMEFT RG evolution}},\ }\href {https://doi.org/10.1103/PhysRevD.108.113005}
  {\bibfield  {journal} {\bibinfo  {journal} {Phys. Rev. D}\ }\textbf {\bibinfo
  {volume} {108}},\ \bibinfo {pages} {113005} (\bibinfo {year} {2023})},\
  \Eprint {https://arxiv.org/abs/2108.05614} {arXiv:2108.05614 [hep-ph]}
  \BibitemShut {NoStop}%
\bibitem [{\citenamefont {Cirigliano}\ \emph
  {et~al.}(2024{\natexlab{a}})\citenamefont {Cirigliano}, \citenamefont
  {Dekens}, \citenamefont {de~Vries}, \citenamefont {Mereghetti},\ and\
  \citenamefont {Tong}}]{Cirigliano:2023nol}%
  \BibitemOpen
  \bibfield  {author} {\bibinfo {author} {\bibfnamefont {V.}~\bibnamefont
  {Cirigliano}}, \bibinfo {author} {\bibfnamefont {W.}~\bibnamefont {Dekens}},
  \bibinfo {author} {\bibfnamefont {J.}~\bibnamefont {de~Vries}}, \bibinfo
  {author} {\bibfnamefont {E.}~\bibnamefont {Mereghetti}},\ and\ \bibinfo
  {author} {\bibfnamefont {T.}~\bibnamefont {Tong}},\ }\bibfield  {title}
  {\bibinfo {title} {{Anomalies in global SMEFT analyses. A case study of
  first-row CKM unitarity}},\ }\href {https://doi.org/10.1007/JHEP03(2024)033}
  {\bibfield  {journal} {\bibinfo  {journal} {JHEP}\ }\textbf {\bibinfo
  {volume} {03}},\ \bibinfo {pages} {033}},\ \Eprint
  {https://arxiv.org/abs/2311.00021} {arXiv:2311.00021 [hep-ph]} \BibitemShut
  {NoStop}%
\bibitem [{\citenamefont {Dawid}\ \emph {et~al.}(2024)\citenamefont {Dawid},
  \citenamefont {Cirigliano},\ and\ \citenamefont {Dekens}}]{Dawid:2024wmp}%
  \BibitemOpen
  \bibfield  {author} {\bibinfo {author} {\bibfnamefont {M.}~\bibnamefont
  {Dawid}}, \bibinfo {author} {\bibfnamefont {V.}~\bibnamefont {Cirigliano}},\
  and\ \bibinfo {author} {\bibfnamefont {W.}~\bibnamefont {Dekens}},\
  }\bibfield  {title} {\bibinfo {title} {{One-loop analysis of
  {\ensuremath{\beta}} decays in SMEFT}},\ }\href
  {https://doi.org/10.1007/JHEP08(2024)175} {\bibfield  {journal} {\bibinfo
  {journal} {JHEP}\ }\textbf {\bibinfo {volume} {08}},\ \bibinfo {pages}
  {175}},\ \Eprint {https://arxiv.org/abs/2402.06723} {arXiv:2402.06723
  [hep-ph]} \BibitemShut {NoStop}%
\bibitem [{\citenamefont {Gorbahn}\ \emph {et~al.}(2023)\citenamefont
  {Gorbahn}, \citenamefont {J\"ager}, \citenamefont {Moretti},\ and\
  \citenamefont {van~der Merwe}}]{Gorbahn:2022rgl}%
  \BibitemOpen
  \bibfield  {author} {\bibinfo {author} {\bibfnamefont {M.}~\bibnamefont
  {Gorbahn}}, \bibinfo {author} {\bibfnamefont {S.}~\bibnamefont {J\"ager}},
  \bibinfo {author} {\bibfnamefont {F.}~\bibnamefont {Moretti}},\ and\ \bibinfo
  {author} {\bibfnamefont {E.}~\bibnamefont {van~der Merwe}},\ }\bibfield
  {title} {\bibinfo {title} {{Semileptonic weak Hamiltonian to $ \mathcal{O}
  $(\ensuremath{\alpha}\ensuremath{\alpha}$_{s}$) in momentum-space subtraction
  schemes}},\ }\href {https://doi.org/10.1007/JHEP01(2023)159} {\bibfield
  {journal} {\bibinfo  {journal} {JHEP}\ }\textbf {\bibinfo {volume} {01}},\
  \bibinfo {pages} {159}},\ \Eprint {https://arxiv.org/abs/2209.05289}
  {arXiv:2209.05289 [hep-ph]} \BibitemShut {NoStop}%
\bibitem [{\citenamefont {Cirigliano}\ \emph
  {et~al.}(2023{\natexlab{a}})\citenamefont {Cirigliano}, \citenamefont
  {Dekens}, \citenamefont {Mereghetti},\ and\ \citenamefont
  {Tomalak}}]{Cirigliano:2023fnz}%
  \BibitemOpen
  \bibfield  {author} {\bibinfo {author} {\bibfnamefont {V.}~\bibnamefont
  {Cirigliano}}, \bibinfo {author} {\bibfnamefont {W.}~\bibnamefont {Dekens}},
  \bibinfo {author} {\bibfnamefont {E.}~\bibnamefont {Mereghetti}},\ and\
  \bibinfo {author} {\bibfnamefont {O.}~\bibnamefont {Tomalak}},\ }\bibfield
  {title} {\bibinfo {title} {{Effective field theory for radiative corrections
  to charged-current processes: Vector coupling}},\ }\href
  {https://doi.org/10.1103/PhysRevD.108.053003} {\bibfield  {journal} {\bibinfo
   {journal} {Phys. Rev. D}\ }\textbf {\bibinfo {volume} {108}},\ \bibinfo
  {pages} {053003} (\bibinfo {year} {2023}{\natexlab{a}})},\ \Eprint
  {https://arxiv.org/abs/2306.03138} {arXiv:2306.03138 [hep-ph]} \BibitemShut
  {NoStop}%
\bibitem [{\citenamefont {Gorbahn}\ \emph {et~al.}(2026)\citenamefont
  {Gorbahn}, \citenamefont {J\"ager}, \citenamefont {Moretti},\ and\
  \citenamefont {van~der Merwe}}]{GJMvM:2026}%
  \BibitemOpen
  \bibfield  {author} {\bibinfo {author} {\bibfnamefont {M.}~\bibnamefont
  {Gorbahn}}, \bibinfo {author} {\bibfnamefont {S.}~\bibnamefont {J\"ager}},
  \bibinfo {author} {\bibfnamefont {F.}~\bibnamefont {Moretti}},\ and\ \bibinfo
  {author} {\bibfnamefont {E.}~\bibnamefont {van~der Merwe}},\ }\href@noop {}
  {\bibinfo {title} {{Higher order correction to the Semi-leptonic Effective
  Operator, in preparation}}} (\bibinfo {year} {2026})\BibitemShut {NoStop}%
\bibitem [{\citenamefont {Hardy}\ and\ \citenamefont
  {Towner}(2020)}]{Hardy:2020qwl}%
  \BibitemOpen
  \bibfield  {author} {\bibinfo {author} {\bibfnamefont {J.~C.}\ \bibnamefont
  {Hardy}}\ and\ \bibinfo {author} {\bibfnamefont {I.~S.}\ \bibnamefont
  {Towner}},\ }\bibfield  {title} {\bibinfo {title} {{Superallowed $0^+ \to
  0^+$ nuclear $\beta$ decays: 2020 critical survey, with implications for
  V$_{ud}$ and CKM unitarity}},\ }\href
  {https://doi.org/10.1103/PhysRevC.102.045501} {\bibfield  {journal} {\bibinfo
   {journal} {Phys. Rev. C}\ }\textbf {\bibinfo {volume} {102}},\ \bibinfo
  {pages} {045501} (\bibinfo {year} {2020})}\BibitemShut {NoStop}%
\bibitem [{\citenamefont {Gorchtein}\ and\ \citenamefont
  {Seng}(2021)}]{Gorchtein:2021fce}%
  \BibitemOpen
  \bibfield  {author} {\bibinfo {author} {\bibfnamefont {M.}~\bibnamefont
  {Gorchtein}}\ and\ \bibinfo {author} {\bibfnamefont {C.-Y.}\ \bibnamefont
  {Seng}},\ }\bibfield  {title} {\bibinfo {title} {{Dispersion relation
  analysis of the radiative corrections to g$_{A}$ in the neutron
  {\ensuremath{\beta}}-decay}},\ }\href
  {https://doi.org/10.1007/JHEP10(2021)053} {\bibfield  {journal} {\bibinfo
  {journal} {JHEP}\ }\textbf {\bibinfo {volume} {10}},\ \bibinfo {pages}
  {053}},\ \Eprint {https://arxiv.org/abs/2106.09185} {arXiv:2106.09185
  [hep-ph]} \BibitemShut {NoStop}%
\bibitem [{\citenamefont {Czarnecki}\ \emph {et~al.}(2019)\citenamefont
  {Czarnecki}, \citenamefont {Marciano},\ and\ \citenamefont
  {Sirlin}}]{Czarnecki:2019mwq}%
  \BibitemOpen
  \bibfield  {author} {\bibinfo {author} {\bibfnamefont {A.}~\bibnamefont
  {Czarnecki}}, \bibinfo {author} {\bibfnamefont {W.~J.}\ \bibnamefont
  {Marciano}},\ and\ \bibinfo {author} {\bibfnamefont {A.}~\bibnamefont
  {Sirlin}},\ }\bibfield  {title} {\bibinfo {title} {{Radiative Corrections to
  Neutron and Nuclear Beta Decays Revisited}},\ }\href
  {https://doi.org/10.1103/PhysRevD.100.073008} {\bibfield  {journal} {\bibinfo
   {journal} {Phys. Rev. D}\ }\textbf {\bibinfo {volume} {100}},\ \bibinfo
  {pages} {073008} (\bibinfo {year} {2019})},\ \Eprint
  {https://arxiv.org/abs/1907.06737} {arXiv:1907.06737 [hep-ph]} \BibitemShut
  {NoStop}%
\bibitem [{\citenamefont {Gorchtein}\ and\ \citenamefont
  {Seng}(2024)}]{Gorchtein:2023naa}%
  \BibitemOpen
  \bibfield  {author} {\bibinfo {author} {\bibfnamefont {M.}~\bibnamefont
  {Gorchtein}}\ and\ \bibinfo {author} {\bibfnamefont {C.~Y.}\ \bibnamefont
  {Seng}},\ }\bibfield  {title} {\bibinfo {title} {{Superallowed Nuclear Beta
  Decays and Precision Tests of the Standard Model}},\ }\href
  {https://doi.org/10.1146/annurev-nucl-102622-020726} {\bibfield  {journal}
  {\bibinfo  {journal} {Ann. Rev. Nucl. Part. Sci.}\ }\textbf {\bibinfo
  {volume} {74}},\ \bibinfo {pages} {23} (\bibinfo {year} {2024})},\ \Eprint
  {https://arxiv.org/abs/2311.00044} {arXiv:2311.00044 [nucl-th]} \BibitemShut
  {NoStop}%
\bibitem [{\citenamefont {Ma}\ \emph {et~al.}(2024)\citenamefont {Ma},
  \citenamefont {Feng}, \citenamefont {Gorchtein}, \citenamefont {Jin},
  \citenamefont {Liu}, \citenamefont {Seng}, \citenamefont {Wang},\ and\
  \citenamefont {Zhang}}]{Ma:2023kfr}%
  \BibitemOpen
  \bibfield  {author} {\bibinfo {author} {\bibfnamefont {P.-X.}\ \bibnamefont
  {Ma}}, \bibinfo {author} {\bibfnamefont {X.}~\bibnamefont {Feng}}, \bibinfo
  {author} {\bibfnamefont {M.}~\bibnamefont {Gorchtein}}, \bibinfo {author}
  {\bibfnamefont {L.-C.}\ \bibnamefont {Jin}}, \bibinfo {author} {\bibfnamefont
  {K.-F.}\ \bibnamefont {Liu}}, \bibinfo {author} {\bibfnamefont {C.-Y.}\
  \bibnamefont {Seng}}, \bibinfo {author} {\bibfnamefont {B.-G.}\ \bibnamefont
  {Wang}},\ and\ \bibinfo {author} {\bibfnamefont {Z.-L.}\ \bibnamefont
  {Zhang}},\ }\bibfield  {title} {\bibinfo {title} {{Lattice QCD Calculation of
  Electroweak Box Contributions to Superallowed Nuclear and Neutron Beta
  Decays}},\ }\href {https://doi.org/10.1103/PhysRevLett.132.191901} {\bibfield
   {journal} {\bibinfo  {journal} {Phys. Rev. Lett.}\ }\textbf {\bibinfo
  {volume} {132}},\ \bibinfo {pages} {191901} (\bibinfo {year} {2024})},\
  \Eprint {https://arxiv.org/abs/2308.16755} {arXiv:2308.16755 [hep-lat]}
  \BibitemShut {NoStop}%
\bibitem [{\citenamefont {Seng}\ \emph {et~al.}(2019)\citenamefont {Seng},
  \citenamefont {Gorchtein},\ and\ \citenamefont
  {Ramsey-Musolf}}]{Seng:2018qru}%
  \BibitemOpen
  \bibfield  {author} {\bibinfo {author} {\bibfnamefont {C.~Y.}\ \bibnamefont
  {Seng}}, \bibinfo {author} {\bibfnamefont {M.}~\bibnamefont {Gorchtein}},\
  and\ \bibinfo {author} {\bibfnamefont {M.~J.}\ \bibnamefont
  {Ramsey-Musolf}},\ }\bibfield  {title} {\bibinfo {title} {{Dispersive
  evaluation of the inner radiative correction in neutron and nuclear $\beta$
  decay}},\ }\href {https://doi.org/10.1103/PhysRevD.100.013001} {\bibfield
  {journal} {\bibinfo  {journal} {Phys. Rev. D}\ }\textbf {\bibinfo {volume}
  {100}},\ \bibinfo {pages} {013001} (\bibinfo {year} {2019})},\ \Eprint
  {https://arxiv.org/abs/1812.03352} {arXiv:1812.03352 [nucl-th]} \BibitemShut
  {NoStop}%
\bibitem [{\citenamefont {Cirigliano}\ \emph
  {et~al.}(2023{\natexlab{b}})\citenamefont {Cirigliano}, \citenamefont
  {Crivellin}, \citenamefont {Hoferichter},\ and\ \citenamefont
  {Moulson}}]{Cirigliano:2022yyo}%
  \BibitemOpen
  \bibfield  {author} {\bibinfo {author} {\bibfnamefont {V.}~\bibnamefont
  {Cirigliano}}, \bibinfo {author} {\bibfnamefont {A.}~\bibnamefont
  {Crivellin}}, \bibinfo {author} {\bibfnamefont {M.}~\bibnamefont
  {Hoferichter}},\ and\ \bibinfo {author} {\bibfnamefont {M.}~\bibnamefont
  {Moulson}},\ }\bibfield  {title} {\bibinfo {title} {{Scrutinizing CKM
  unitarity with a new measurement of the \ensuremath{{\rm K}_{\mu
  3}}/\ensuremath{{\rm K}_{\mu 2}} branching fraction}},\ }\href
  {https://doi.org/10.1016/j.physletb.2023.137748} {\bibfield  {journal}
  {\bibinfo  {journal} {Phys. Lett. B}\ }\textbf {\bibinfo {volume} {838}},\
  \bibinfo {pages} {137748} (\bibinfo {year} {2023}{\natexlab{b}})},\ \Eprint
  {https://arxiv.org/abs/2208.11707} {arXiv:2208.11707 [hep-ph]} \BibitemShut
  {NoStop}%
\bibitem [{\citenamefont {Marciano}\ and\ \citenamefont
  {Sirlin}(2006)}]{Marciano:2005ec}%
  \BibitemOpen
  \bibfield  {author} {\bibinfo {author} {\bibfnamefont {W.~J.}\ \bibnamefont
  {Marciano}}\ and\ \bibinfo {author} {\bibfnamefont {A.}~\bibnamefont
  {Sirlin}},\ }\bibfield  {title} {\bibinfo {title} {{Improved calculation of
  electroweak radiative corrections and the value of V(ud)}},\ }\href
  {https://doi.org/10.1103/PhysRevLett.96.032002} {\bibfield  {journal}
  {\bibinfo  {journal} {Phys. Rev. Lett.}\ }\textbf {\bibinfo {volume} {96}},\
  \bibinfo {pages} {032002} (\bibinfo {year} {2006})},\ \Eprint
  {https://arxiv.org/abs/hep-ph/0510099} {arXiv:hep-ph/0510099} \BibitemShut
  {NoStop}%
\bibitem [{\citenamefont {Seng}\ \emph {et~al.}(2018)\citenamefont {Seng},
  \citenamefont {Gorchtein}, \citenamefont {Patel},\ and\ \citenamefont
  {Ramsey-Musolf}}]{Seng:2018yzq}%
  \BibitemOpen
  \bibfield  {author} {\bibinfo {author} {\bibfnamefont {C.-Y.}\ \bibnamefont
  {Seng}}, \bibinfo {author} {\bibfnamefont {M.}~\bibnamefont {Gorchtein}},
  \bibinfo {author} {\bibfnamefont {H.~H.}\ \bibnamefont {Patel}},\ and\
  \bibinfo {author} {\bibfnamefont {M.~J.}\ \bibnamefont {Ramsey-Musolf}},\
  }\bibfield  {title} {\bibinfo {title} {{Reduced Hadronic Uncertainty in the
  Determination of $V_{ud}$}},\ }\href
  {https://doi.org/10.1103/PhysRevLett.121.241804} {\bibfield  {journal}
  {\bibinfo  {journal} {Phys. Rev. Lett.}\ }\textbf {\bibinfo {volume} {121}},\
  \bibinfo {pages} {241804} (\bibinfo {year} {2018})},\ \Eprint
  {https://arxiv.org/abs/1807.10197} {arXiv:1807.10197 [hep-ph]} \BibitemShut
  {NoStop}%
\bibitem [{\citenamefont {Seng}\ and\ \citenamefont
  {Gorchtein}(2023)}]{Seng:2022cnq}%
  \BibitemOpen
  \bibfield  {author} {\bibinfo {author} {\bibfnamefont {C.-Y.}\ \bibnamefont
  {Seng}}\ and\ \bibinfo {author} {\bibfnamefont {M.}~\bibnamefont
  {Gorchtein}},\ }\bibfield  {title} {\bibinfo {title} {{Dispersive formalism
  for the nuclear structure correction {\ensuremath{\delta}}NS to the
  {\ensuremath{\beta}} decay rate}},\ }\href
  {https://doi.org/10.1103/PhysRevC.107.035503} {\bibfield  {journal} {\bibinfo
   {journal} {Phys. Rev. C}\ }\textbf {\bibinfo {volume} {107}},\ \bibinfo
  {pages} {035503} (\bibinfo {year} {2023})},\ \Eprint
  {https://arxiv.org/abs/2211.10214} {arXiv:2211.10214 [nucl-th]} \BibitemShut
  {NoStop}%
\bibitem [{\citenamefont {Towner}\ and\ \citenamefont
  {Hardy}(2010)}]{Towner:2010zz}%
  \BibitemOpen
  \bibfield  {author} {\bibinfo {author} {\bibfnamefont {I.~S.}\ \bibnamefont
  {Towner}}\ and\ \bibinfo {author} {\bibfnamefont {J.~C.}\ \bibnamefont
  {Hardy}},\ }\bibfield  {title} {\bibinfo {title} {{The evaluation of V(ud)
  and its impact on the unitarity of the Cabibbo-Kobayashi-Maskawa quark-mixing
  matrix}},\ }\href {https://doi.org/10.1088/0034-4885/73/4/046301} {\bibfield
  {journal} {\bibinfo  {journal} {Rept. Prog. Phys.}\ }\textbf {\bibinfo
  {volume} {73}},\ \bibinfo {pages} {046301} (\bibinfo {year}
  {2010})}\BibitemShut {NoStop}%
\bibitem [{\citenamefont {Czarnecki}\ \emph {et~al.}(2004)\citenamefont
  {Czarnecki}, \citenamefont {Marciano},\ and\ \citenamefont
  {Sirlin}}]{Czarnecki:2004cw}%
  \BibitemOpen
  \bibfield  {author} {\bibinfo {author} {\bibfnamefont {A.}~\bibnamefont
  {Czarnecki}}, \bibinfo {author} {\bibfnamefont {W.~J.}\ \bibnamefont
  {Marciano}},\ and\ \bibinfo {author} {\bibfnamefont {A.}~\bibnamefont
  {Sirlin}},\ }\bibfield  {title} {\bibinfo {title} {{Precision measurements
  and CKM unitarity}},\ }\href {https://doi.org/10.1103/PhysRevD.70.093006}
  {\bibfield  {journal} {\bibinfo  {journal} {Phys. Rev. D}\ }\textbf {\bibinfo
  {volume} {70}},\ \bibinfo {pages} {093006} (\bibinfo {year} {2004})},\
  \Eprint {https://arxiv.org/abs/hep-ph/0406324} {arXiv:hep-ph/0406324}
  \BibitemShut {NoStop}%
\bibitem [{\citenamefont {Cirigliano}\ \emph
  {et~al.}(2024{\natexlab{b}})\citenamefont {Cirigliano}, \citenamefont
  {Dekens}, \citenamefont {de~Vries}, \citenamefont {Gandolfi}, \citenamefont
  {Hoferichter},\ and\ \citenamefont {Mereghetti}}]{Cirigliano:2024msg}%
  \BibitemOpen
  \bibfield  {author} {\bibinfo {author} {\bibfnamefont {V.}~\bibnamefont
  {Cirigliano}}, \bibinfo {author} {\bibfnamefont {W.}~\bibnamefont {Dekens}},
  \bibinfo {author} {\bibfnamefont {J.}~\bibnamefont {de~Vries}}, \bibinfo
  {author} {\bibfnamefont {S.}~\bibnamefont {Gandolfi}}, \bibinfo {author}
  {\bibfnamefont {M.}~\bibnamefont {Hoferichter}},\ and\ \bibinfo {author}
  {\bibfnamefont {E.}~\bibnamefont {Mereghetti}},\ }\bibfield  {title}
  {\bibinfo {title} {{Ab initio electroweak corrections to superallowed
  {\ensuremath{\beta}} decays and their impact on Vud}},\ }\href
  {https://doi.org/10.1103/PhysRevC.110.055502} {\bibfield  {journal} {\bibinfo
   {journal} {Phys. Rev. C}\ }\textbf {\bibinfo {volume} {110}},\ \bibinfo
  {pages} {055502} (\bibinfo {year} {2024}{\natexlab{b}})},\ \Eprint
  {https://arxiv.org/abs/2405.18464} {arXiv:2405.18464 [nucl-th]} \BibitemShut
  {NoStop}%
\bibitem [{\citenamefont {Cirigliano}\ \emph
  {et~al.}(2024{\natexlab{c}})\citenamefont {Cirigliano}, \citenamefont
  {Dekens}, \citenamefont {de~Vries}, \citenamefont {Gandolfi}, \citenamefont
  {Hoferichter},\ and\ \citenamefont {Mereghetti}}]{Cirigliano:2024rfk}%
  \BibitemOpen
  \bibfield  {author} {\bibinfo {author} {\bibfnamefont {V.}~\bibnamefont
  {Cirigliano}}, \bibinfo {author} {\bibfnamefont {W.}~\bibnamefont {Dekens}},
  \bibinfo {author} {\bibfnamefont {J.}~\bibnamefont {de~Vries}}, \bibinfo
  {author} {\bibfnamefont {S.}~\bibnamefont {Gandolfi}}, \bibinfo {author}
  {\bibfnamefont {M.}~\bibnamefont {Hoferichter}},\ and\ \bibinfo {author}
  {\bibfnamefont {E.}~\bibnamefont {Mereghetti}},\ }\bibfield  {title}
  {\bibinfo {title} {{Radiative Corrections to Superallowed
  {\ensuremath{\beta}} Decays in Effective Field Theory}},\ }\href
  {https://doi.org/10.1103/PhysRevLett.133.211801} {\bibfield  {journal}
  {\bibinfo  {journal} {Phys. Rev. Lett.}\ }\textbf {\bibinfo {volume} {133}},\
  \bibinfo {pages} {211801} (\bibinfo {year} {2024}{\natexlab{c}})},\ \Eprint
  {https://arxiv.org/abs/2405.18469} {arXiv:2405.18469 [hep-ph]} \BibitemShut
  {NoStop}%
\bibitem [{\citenamefont {Ando}\ \emph {et~al.}(2004)\citenamefont {Ando},
  \citenamefont {Fearing}, \citenamefont {Gudkov}, \citenamefont {Kubodera},
  \citenamefont {Myhrer}, \citenamefont {Nakamura},\ and\ \citenamefont
  {Sato}}]{Ando:2004rk}%
  \BibitemOpen
  \bibfield  {author} {\bibinfo {author} {\bibfnamefont {S.}~\bibnamefont
  {Ando}}, \bibinfo {author} {\bibfnamefont {H.~W.}\ \bibnamefont {Fearing}},
  \bibinfo {author} {\bibfnamefont {V.~P.}\ \bibnamefont {Gudkov}}, \bibinfo
  {author} {\bibfnamefont {K.}~\bibnamefont {Kubodera}}, \bibinfo {author}
  {\bibfnamefont {F.}~\bibnamefont {Myhrer}}, \bibinfo {author} {\bibfnamefont
  {S.}~\bibnamefont {Nakamura}},\ and\ \bibinfo {author} {\bibfnamefont
  {T.}~\bibnamefont {Sato}},\ }\bibfield  {title} {\bibinfo {title} {{Neutron
  beta decay in effective field theory}},\ }\href
  {https://doi.org/10.1016/j.physletb.2004.06.037} {\bibfield  {journal}
  {\bibinfo  {journal} {Phys. Lett. B}\ }\textbf {\bibinfo {volume} {595}},\
  \bibinfo {pages} {250} (\bibinfo {year} {2004})},\ \Eprint
  {https://arxiv.org/abs/nucl-th/0402100} {arXiv:nucl-th/0402100} \BibitemShut
  {NoStop}%
\bibitem [{\citenamefont {Raha}\ \emph {et~al.}(2012)\citenamefont {Raha},
  \citenamefont {Myhrer},\ and\ \citenamefont {Kubodera}}]{Raha:2011aa}%
  \BibitemOpen
  \bibfield  {author} {\bibinfo {author} {\bibfnamefont {U.}~\bibnamefont
  {Raha}}, \bibinfo {author} {\bibfnamefont {F.}~\bibnamefont {Myhrer}},\ and\
  \bibinfo {author} {\bibfnamefont {K.}~\bibnamefont {Kubodera}},\ }\bibfield
  {title} {\bibinfo {title} {{Radiative corrections to anti-neutrino proton
  scattering at low energies}},\ }\href
  {https://doi.org/10.1103/PhysRevC.85.045502} {\bibfield  {journal} {\bibinfo
  {journal} {Phys. Rev. C}\ }\textbf {\bibinfo {volume} {85}},\ \bibinfo
  {pages} {045502} (\bibinfo {year} {2012})},\ \bibinfo {note} {[Erratum:
  Phys.Rev.C 86, 039903 (2012)]},\ \Eprint {https://arxiv.org/abs/1112.2007}
  {arXiv:1112.2007 [hep-ph]} \BibitemShut {NoStop}%
\bibitem [{\citenamefont {Falkowski}\ \emph {et~al.}(2024)\citenamefont
  {Falkowski}, \citenamefont {Gonzalez-Alonso}, \citenamefont {Palavric},\ and\
  \citenamefont {Rodrguez-Sanchez}}]{Falkowski:2021vdg}%
  \BibitemOpen
  \bibfield  {author} {\bibinfo {author} {\bibfnamefont {A.}~\bibnamefont
  {Falkowski}}, \bibinfo {author} {\bibfnamefont {M.}~\bibnamefont
  {Gonzalez-Alonso}}, \bibinfo {author} {\bibfnamefont {A.}~\bibnamefont
  {Palavric}},\ and\ \bibinfo {author} {\bibfnamefont {A.}~\bibnamefont
  {Rodrguez-Sanchez}},\ }\bibfield  {title} {\bibinfo {title} {{Constraints on
  subleading interactions in beta decay Lagrangian}},\ }\href
  {https://doi.org/10.1007/JHEP02(2024)091} {\bibfield  {journal} {\bibinfo
  {journal} {JHEP}\ }\textbf {\bibinfo {volume} {02}},\ \bibinfo {pages}
  {091}},\ \Eprint {https://arxiv.org/abs/2112.07688} {arXiv:2112.07688
  [hep-ph]} \BibitemShut {NoStop}%
\bibitem [{\citenamefont {Descotes-Genon}\ and\ \citenamefont
  {Moussallam}(2005)}]{Descotes-Genon:2005wrq}%
  \BibitemOpen
  \bibfield  {author} {\bibinfo {author} {\bibfnamefont {S.}~\bibnamefont
  {Descotes-Genon}}\ and\ \bibinfo {author} {\bibfnamefont {B.}~\bibnamefont
  {Moussallam}},\ }\bibfield  {title} {\bibinfo {title} {{Radiative corrections
  in weak semi-leptonic processes at low energy: A Two-step matching
  determination}},\ }\href {https://doi.org/10.1140/epjc/s2005-02316-8}
  {\bibfield  {journal} {\bibinfo  {journal} {Eur. Phys. J. C}\ }\textbf
  {\bibinfo {volume} {42}},\ \bibinfo {pages} {403} (\bibinfo {year} {2005})},\
  \Eprint {https://arxiv.org/abs/hep-ph/0505077} {arXiv:hep-ph/0505077}
  \BibitemShut {NoStop}%
\bibitem [{Note1()}]{Note1}%
  \BibitemOpen
  \bibinfo {note} {Our definition \protect \textup {\hbox {\mathsurround \z@
  \protect \normalfont (\ignorespaces \ref {eq:evanescent-operator}\unskip
  \@@italiccorr )}} is equivalent to the one in Ref.~\cite {Cirigliano:2023fnz}
  after identifying $a_{\protect \rm Ref.~\cite {Cirigliano:2023fnz}}\to
  (a-3)/2$.}\BibitemShut {Stop}%
\bibitem [{Note2()}]{Note2}%
  \BibitemOpen
  \bibinfo {note} {We confirmed at two loop that the subtraction term exactly
  reproduced the weak effective theory using different IR regulators such as a
  photon mass or off-shell momenta.}\BibitemShut {Stop}%
\bibitem [{\citenamefont {Seng}\ \emph {et~al.}(2020)\citenamefont {Seng},
  \citenamefont {Feng}, \citenamefont {Gorchtein},\ and\ \citenamefont
  {Jin}}]{Seng:2020wjq}%
  \BibitemOpen
  \bibfield  {author} {\bibinfo {author} {\bibfnamefont {C.-Y.}\ \bibnamefont
  {Seng}}, \bibinfo {author} {\bibfnamefont {X.}~\bibnamefont {Feng}}, \bibinfo
  {author} {\bibfnamefont {M.}~\bibnamefont {Gorchtein}},\ and\ \bibinfo
  {author} {\bibfnamefont {L.-C.}\ \bibnamefont {Jin}},\ }\bibfield  {title}
  {\bibinfo {title} {{Joint lattice QCD{\textendash}dispersion theory analysis
  confirms the quark-mixing top-row unitarity deficit}},\ }\href
  {https://doi.org/10.1103/PhysRevD.101.111301} {\bibfield  {journal} {\bibinfo
   {journal} {Phys. Rev. D}\ }\textbf {\bibinfo {volume} {101}},\ \bibinfo
  {pages} {111301} (\bibinfo {year} {2020})},\ \Eprint
  {https://arxiv.org/abs/2003.11264} {arXiv:2003.11264 [hep-ph]} \BibitemShut
  {NoStop}%
\bibitem [{\citenamefont {Hayen}(2021)}]{Hayen:2020cxh}%
  \BibitemOpen
  \bibfield  {author} {\bibinfo {author} {\bibfnamefont {L.}~\bibnamefont
  {Hayen}},\ }\bibfield  {title} {\bibinfo {title} {{Standard model
  $\mathcal{O}(\alpha)$ renormalization of $g_A$ and its impact on new physics
  searches}},\ }\href {https://doi.org/10.1103/PhysRevD.103.113001} {\bibfield
  {journal} {\bibinfo  {journal} {Phys. Rev. D}\ }\textbf {\bibinfo {volume}
  {103}},\ \bibinfo {pages} {113001} (\bibinfo {year} {2021})},\ \Eprint
  {https://arxiv.org/abs/2010.07262} {arXiv:2010.07262 [hep-ph]} \BibitemShut
  {NoStop}%
\bibitem [{\citenamefont {Shiells}\ \emph {et~al.}(2021)\citenamefont
  {Shiells}, \citenamefont {Blunden},\ and\ \citenamefont
  {Melnitchouk}}]{Shiells:2020fqp}%
  \BibitemOpen
  \bibfield  {author} {\bibinfo {author} {\bibfnamefont {K.}~\bibnamefont
  {Shiells}}, \bibinfo {author} {\bibfnamefont {P.~G.}\ \bibnamefont
  {Blunden}},\ and\ \bibinfo {author} {\bibfnamefont {W.}~\bibnamefont
  {Melnitchouk}},\ }\bibfield  {title} {\bibinfo {title} {{Electroweak axial
  structure functions and improved extraction of the Vud CKM matrix element}},\
  }\href {https://doi.org/10.1103/PhysRevD.104.033003} {\bibfield  {journal}
  {\bibinfo  {journal} {Phys. Rev. D}\ }\textbf {\bibinfo {volume} {104}},\
  \bibinfo {pages} {033003} (\bibinfo {year} {2021})},\ \Eprint
  {https://arxiv.org/abs/2012.01580} {arXiv:2012.01580 [hep-ph]} \BibitemShut
  {NoStop}%
\bibitem [{\citenamefont {Larin}\ \emph {et~al.}(1991)\citenamefont {Larin},
  \citenamefont {Tkachov},\ and\ \citenamefont {Vermaseren}}]{Larin:1990zw}%
  \BibitemOpen
  \bibfield  {author} {\bibinfo {author} {\bibfnamefont {S.~A.}\ \bibnamefont
  {Larin}}, \bibinfo {author} {\bibfnamefont {F.~V.}\ \bibnamefont {Tkachov}},\
  and\ \bibinfo {author} {\bibfnamefont {J.~A.~M.}\ \bibnamefont
  {Vermaseren}},\ }\bibfield  {title} {\bibinfo {title} {{The $\alpha_s^3$
  correction to the Bjorken sum rule}},\ }\href
  {https://doi.org/10.1103/PhysRevLett.66.862} {\bibfield  {journal} {\bibinfo
  {journal} {Phys. Rev. Lett.}\ }\textbf {\bibinfo {volume} {66}},\ \bibinfo
  {pages} {862} (\bibinfo {year} {1991})}\BibitemShut {NoStop}%
\bibitem [{\citenamefont {Larin}\ and\ \citenamefont
  {Vermaseren}(1991)}]{Larin:1991tj}%
  \BibitemOpen
  \bibfield  {author} {\bibinfo {author} {\bibfnamefont {S.~A.}\ \bibnamefont
  {Larin}}\ and\ \bibinfo {author} {\bibfnamefont {J.~A.~M.}\ \bibnamefont
  {Vermaseren}},\ }\bibfield  {title} {\bibinfo {title} {{The
  $\alpha_\mathrm{s}^3$ corrections to the Bjorken sum rule for polarized
  electroproduction and to the Gross-Llewellyn Smith sum rule}},\ }\href
  {https://doi.org/10.1016/0370-2693(91)90839-I} {\bibfield  {journal}
  {\bibinfo  {journal} {Phys. Lett. B}\ }\textbf {\bibinfo {volume} {259}},\
  \bibinfo {pages} {345} (\bibinfo {year} {1991})}\BibitemShut {NoStop}%
\bibitem [{\citenamefont {Baikov}\ \emph {et~al.}(2010)\citenamefont {Baikov},
  \citenamefont {Chetyrkin},\ and\ \citenamefont {Kuhn}}]{Baikov:2010je}%
  \BibitemOpen
  \bibfield  {author} {\bibinfo {author} {\bibfnamefont {P.~A.}\ \bibnamefont
  {Baikov}}, \bibinfo {author} {\bibfnamefont {K.~G.}\ \bibnamefont
  {Chetyrkin}},\ and\ \bibinfo {author} {\bibfnamefont {J.~H.}\ \bibnamefont
  {Kuhn}},\ }\bibfield  {title} {\bibinfo {title} {{Adler Function, Bjorken Sum
  Rule, and the Crewther Relation to Order $\alpha^4_s$ in a General Gauge
  Theory}},\ }\href {https://doi.org/10.1103/PhysRevLett.104.132004} {\bibfield
   {journal} {\bibinfo  {journal} {Phys. Rev. Lett.}\ }\textbf {\bibinfo
  {volume} {104}},\ \bibinfo {pages} {132004} (\bibinfo {year} {2010})},\
  \Eprint {https://arxiv.org/abs/1001.3606} {arXiv:1001.3606 [hep-ph]}
  \BibitemShut {NoStop}%
\bibitem [{Note3()}]{Note3}%
  \BibitemOpen
  \bibinfo {note} {We take into account the $\tau $ threshold in the running of
  $\alpha $, but we decouple the $\tau $ and the charm at the same scale in the
  running of the Wilson coefficient. We checked that the effect of decoupling
  the $\tau $ at $m_{\tau }$ in the running of the Wilson coefficient is
  negligible.}\BibitemShut {Stop}%
\bibitem [{\citenamefont {Gambino}\ and\ \citenamefont
  {Haisch}(2001)}]{Gambino:2001au}%
  \BibitemOpen
  \bibfield  {author} {\bibinfo {author} {\bibfnamefont {P.}~\bibnamefont
  {Gambino}}\ and\ \bibinfo {author} {\bibfnamefont {U.}~\bibnamefont
  {Haisch}},\ }\bibfield  {title} {\bibinfo {title} {{Complete electroweak
  matching for radiative B decays}},\ }\href
  {https://doi.org/10.1088/1126-6708/2001/10/020} {\bibfield  {journal}
  {\bibinfo  {journal} {JHEP}\ }\textbf {\bibinfo {volume} {10}},\ \bibinfo
  {pages} {020}},\ \Eprint {https://arxiv.org/abs/hep-ph/0109058}
  {arXiv:hep-ph/0109058} \BibitemShut {NoStop}%
\bibitem [{\citenamefont {Brod}\ and\ \citenamefont
  {Gorbahn}(2008)}]{Brod:2008ss}%
  \BibitemOpen
  \bibfield  {author} {\bibinfo {author} {\bibfnamefont {J.}~\bibnamefont
  {Brod}}\ and\ \bibinfo {author} {\bibfnamefont {M.}~\bibnamefont {Gorbahn}},\
  }\bibfield  {title} {\bibinfo {title} {{Electroweak Corrections to the Charm
  Quark Contribution to $K^+ \to \pi^+ \nu \bar{\nu}$}},\ }\href
  {https://doi.org/10.1103/PhysRevD.78.034006} {\bibfield  {journal} {\bibinfo
  {journal} {Phys. Rev. D}\ }\textbf {\bibinfo {volume} {78}},\ \bibinfo
  {pages} {034006} (\bibinfo {year} {2008})},\ \Eprint
  {https://arxiv.org/abs/hep-ph/0805.4119} {arXiv:hep-ph/0805.4119 [hep-ph]}
  \BibitemShut {NoStop}%
\bibitem [{\citenamefont {Borah}\ \emph {et~al.}(2024)\citenamefont {Borah},
  \citenamefont {Hill},\ and\ \citenamefont {Plestid}}]{Borah:2024ghn}%
  \BibitemOpen
  \bibfield  {author} {\bibinfo {author} {\bibfnamefont {K.}~\bibnamefont
  {Borah}}, \bibinfo {author} {\bibfnamefont {R.~J.}\ \bibnamefont {Hill}},\
  and\ \bibinfo {author} {\bibfnamefont {R.}~\bibnamefont {Plestid}},\
  }\bibfield  {title} {\bibinfo {title} {{Renormalization of beta decay at
  three loops and beyond}},\ }\href
  {https://doi.org/10.1103/PhysRevD.109.113007} {\bibfield  {journal} {\bibinfo
   {journal} {Phys. Rev. D}\ }\textbf {\bibinfo {volume} {109}},\ \bibinfo
  {pages} {113007} (\bibinfo {year} {2024})},\ \Eprint
  {https://arxiv.org/abs/2402.13307} {arXiv:2402.13307 [hep-ph]} \BibitemShut
  {NoStop}%
\bibitem [{\citenamefont {Hill}\ and\ \citenamefont
  {Plestid}(2024{\natexlab{a}})}]{Hill:2023acw}%
  \BibitemOpen
  \bibfield  {author} {\bibinfo {author} {\bibfnamefont {R.~J.}\ \bibnamefont
  {Hill}}\ and\ \bibinfo {author} {\bibfnamefont {R.}~\bibnamefont {Plestid}},\
  }\bibfield  {title} {\bibinfo {title} {{Field Theory of the Fermi
  Function}},\ }\href {https://doi.org/10.1103/PhysRevLett.133.021803}
  {\bibfield  {journal} {\bibinfo  {journal} {Phys. Rev. Lett.}\ }\textbf
  {\bibinfo {volume} {133}},\ \bibinfo {pages} {021803} (\bibinfo {year}
  {2024}{\natexlab{a}})},\ \Eprint {https://arxiv.org/abs/2309.07343}
  {arXiv:2309.07343 [hep-ph]} \BibitemShut {NoStop}%
\bibitem [{\citenamefont {Hill}\ and\ \citenamefont
  {Plestid}(2024{\natexlab{b}})}]{Hill:2023bfh}%
  \BibitemOpen
  \bibfield  {author} {\bibinfo {author} {\bibfnamefont {R.~J.}\ \bibnamefont
  {Hill}}\ and\ \bibinfo {author} {\bibfnamefont {R.}~\bibnamefont {Plestid}},\
  }\bibfield  {title} {\bibinfo {title} {{All orders factorization and the
  Coulomb problem}},\ }\href {https://doi.org/10.1103/PhysRevD.109.056006}
  {\bibfield  {journal} {\bibinfo  {journal} {Phys. Rev. D}\ }\textbf {\bibinfo
  {volume} {109}},\ \bibinfo {pages} {056006} (\bibinfo {year}
  {2024}{\natexlab{b}})},\ \Eprint {https://arxiv.org/abs/2309.15929}
  {arXiv:2309.15929 [hep-ph]} \BibitemShut {NoStop}%
\bibitem [{\citenamefont {Plestid}(2024)}]{Plestid:2024eib}%
  \BibitemOpen
  \bibfield  {author} {\bibinfo {author} {\bibfnamefont {R.}~\bibnamefont
  {Plestid}},\ }\bibfield  {title} {\bibinfo {title} {{Generalized eikonal
  identities for charged currents}},\ }\href
  {https://doi.org/10.1007/JHEP07(2024)216} {\bibfield  {journal} {\bibinfo
  {journal} {JHEP}\ }\textbf {\bibinfo {volume} {07}},\ \bibinfo {pages}
  {216}},\ \Eprint {https://arxiv.org/abs/2402.14769} {arXiv:2402.14769
  [hep-ph]} \BibitemShut {NoStop}%
\bibitem [{\citenamefont {Vander~Griend}\ \emph {et~al.}(2025)\citenamefont
  {Vander~Griend}, \citenamefont {Cao}, \citenamefont {Hill},\ and\
  \citenamefont {Plestid}}]{VanderGriend:2025mdc}%
  \BibitemOpen
  \bibfield  {author} {\bibinfo {author} {\bibfnamefont {P.}~\bibnamefont
  {Vander~Griend}}, \bibinfo {author} {\bibfnamefont {Z.}~\bibnamefont {Cao}},
  \bibinfo {author} {\bibfnamefont {R.~J.}\ \bibnamefont {Hill}},\ and\
  \bibinfo {author} {\bibfnamefont {R.}~\bibnamefont {Plestid}},\ }\bibfield
  {title} {\bibinfo {title} {{The Fermi function and the neutron's lifetime}},\
  }\href {https://doi.org/10.1016/j.physletb.2025.139678} {\bibfield  {journal}
  {\bibinfo  {journal} {Phys. Lett. B}\ }\textbf {\bibinfo {volume} {868}},\
  \bibinfo {pages} {139678} (\bibinfo {year} {2025})},\ \Eprint
  {https://arxiv.org/abs/2501.17916} {arXiv:2501.17916 [hep-ph]} \BibitemShut
  {NoStop}%
\bibitem [{\citenamefont {Cao}\ \emph {et~al.}(2025)\citenamefont {Cao},
  \citenamefont {Hill}, \citenamefont {Plestid},\ and\ \citenamefont
  {Vander~Griend}}]{Cao:2025lrw}%
  \BibitemOpen
  \bibfield  {author} {\bibinfo {author} {\bibfnamefont {Z.}~\bibnamefont
  {Cao}}, \bibinfo {author} {\bibfnamefont {R.~J.}\ \bibnamefont {Hill}},
  \bibinfo {author} {\bibfnamefont {R.}~\bibnamefont {Plestid}},\ and\ \bibinfo
  {author} {\bibfnamefont {P.}~\bibnamefont {Vander~Griend}},\ }\bibfield
  {title} {\bibinfo {title} {{Factorization and resummation of QED radiative
  corrections for neutron beta decay}},\ }\href@noop {} {\  (\bibinfo {year}
  {2025})},\ \Eprint {https://arxiv.org/abs/2508.05741} {arXiv:2508.05741
  [hep-ph]} \BibitemShut {NoStop}%
\bibitem [{Note4()}]{Note4}%
  \BibitemOpen
  \bibinfo {note} {Using the nonperturbative input $\Box ^{V<}_{\gamma
  W}(Q_0^2) = 1.560(110)(10)(48)\cdot 10^{-3}$ from Ref.~\cite
  {Cirigliano:2023fnz}, which sums the Regge, resonance, and elastic
  contribution with errors in brackets, we find $(g_{V} - 1)_{\protect \text
  {LL}_s} \cdot 10^6 = 11478^{+60}_{-57}$, corresponding to a central value of
  $\Delta ^V_R = 2.461\%$, in agreement with their value within perturbative
  uncertainties.}\BibitemShut {Stop}%
\bibitem [{\citenamefont {Vermaseren}(1994)}]{Vermaseren:1994je}%
  \BibitemOpen
  \bibfield  {author} {\bibinfo {author} {\bibfnamefont {J.~A.~M.}\
  \bibnamefont {Vermaseren}},\ }\bibfield  {title} {\bibinfo {title}
  {{Axodraw}},\ }\href {https://doi.org/10.1016/0010-4655(94)90034-5}
  {\bibfield  {journal} {\bibinfo  {journal} {Comput. Phys. Commun.}\ }\textbf
  {\bibinfo {volume} {83}},\ \bibinfo {pages} {45} (\bibinfo {year}
  {1994})}\BibitemShut {NoStop}%
\end{thebibliography}%

\end{document}